\def\N {I\!\!N}
\def\H {I\!\!H}
\def\R {I\!\!R}
\def\Om{\Omega}
\def\zZ{{\hbox{$\textstyle Z\hskip -5pt Z$}}}
\def\al{\alpha}

\documentclass[12pt]{article}

\usepackage{graphicx}
\newtheorem{Theorem}{Theorem}[section]
\newtheorem{Lemma}{Lemma}[section]

\def\ve{\varepsilon}

\begin{document}
\begin{center}

%{\bf \large Multifractal spectrum and Thermodynamical formalism of
%the Farey Tree }

{\bf \large MULTIFRACTAL SPECTRUM AND THERMODYNAMICAL FORMALISM OF THE FAREY TREE}

\bigskip

 M. PIACQUADIO and E. CESARATTO

{\small \it Departmento
de Matem\'atica, Facultad de Ingenier\'\i a, 

Universidad de Buenos
Aires, Avenida Paseo Colon 850,

 (1063) Buenos Aires, Argentina.
email: ecesara@tron.fi.ar}
 
\end{center}

\abstract

Let $(\Omega, \mu)$ be a set of real numbers to which we associate
a measure $\mu$. Let $\al \ge 0$, let $\Omega_{\al} = \{ x \in
\Omega / \al(x) = \al \},$ where $\al$ is the concentration index
defined by Halsey et al. [Halsey et al., 1986]. Let $f_H(\al)$ be the Hausdorff
dimension of $\Omega_{\al}$. Let $f_L(\al)$ be the Legendre
spectrum of $\Omega$, as defined in [Riedi and Mandelbrot, 1998]; and $f_C(\al)$ the
classical computational spectrum of $\Omega$, defined in [Halsey et al., 1986]. The
task of comparing $f_H, f_C,$ and $f_L$ for different measures
$\mu$ was tackled by several authors ([Cawley and Mauldin, 1992], [Mandelbrot and Riedi, 1997], [Riedi and Mandelbrot, 1998]) working,
mainly, on self similar measures $\mu$. The Farey tree partition
in the unit segment induces a probability measure $\mu$ on an
universal class of fractal sets $\Omega$ that occur in Physics and
other disciplines. This measure $\mu$ is the Hyperbolic measure
$\mu^{\H}$, fundamentally different from any self-similar one. In
this paper we compare $f_H, f_C, $ and $f_L$ for $\mu^{\H}$.

\section{Introduction}

In recent papers [Cawley and Mauldin, 1992], [Mandelbrot and Riedi, 1997], [Riedi and Mandelbrot, 1998], a number of multifractal spectra
$(\al, f(\al))$ of a certain object $\Om$ were defined and their various properties studied. For instance $(\al , f_G(\al))$ [Riedi and Mandelbrot, 1998]
is the mathematically correct version of the original spectrum [Halsey et al., 1986]
defined in 1986 by Halsey, Procaccia and others. But there are
other spectra, such as $f_H(\al)$, which is the Hausdorff
dimension of the set of all elements in $\Om$ sharing the same
concentration $\al$; furthermore, we have spectra like $f_p$
[Riedi and Mandelbrot, 1998],... and so on.

Throughout this paper, $\Om$ will be the unit segment. For the
so-called "self-similar" measures of $\Om$, "all reasonably well defined
spectra coincide, and fulfill the Legendre equations" [Riedi and Mandelbrot, 1998].

\subsection{The original spectrum  $f(\al)$}

\medskip
Let us introduce this spectrum [Halsey et al., 1986] with a very well known example that will be
relevant for our work below. Let $\mu$ be a probability measure on
$\Om$, given, say, in an experimental way: we are measuring,
empirically, a certain variable that we will name $x$, $x \in
[0,1]$. We start the experiment from an initial value of $x$,
namely $x_0$, and we determine another value of $x$, say $x_1 \in
(0,1)$. We will consider intervals $[0,x_1]$ and $[x_1,1]$ as
equiprobable, i.e. we will write $\mu([0,x_1]) = \mu([x_1, 1]) =
\frac{1}{2^1}$. We repeat the experiment, now {\it taking the value {$x_1$} as
the initial value}, and we obtain
another value of $x$, say $x_2$, and, after another iteration of
the process |$x_2$ the new initial value| we obtain $x_3$. Suppose
that we find, e.g., $x_2 \in (x_1, 1)$, whereas $x_3 \in (0,
x_1)$. Then, we state: $\mu([0,x_3]) = \mu([x_3,x_1]) = \mu([x_1,
x_2]) = \mu([x_2, 1]) = \frac{1}{4} = \frac{1}{2^2}$; i.e. we
defined $\mu$ to be equiprobable on the 4 segments of the second
partition of the unit segment. Next, let us suppose that the
reiteration of the process provides us with an $x_4$,  $x_5$,
$x_6$, and $x_7$ inside the interior of the four segments above,
then we say that we have the third partition of $[0,1]$, we obtain $8 =
2^3$ segments, unequal in length, and we define $\mu$ to be
equiprobable over all eight of them, i.e. $\mu$ will be $1/2^3$
for each segment in the third partition. Partitions obtained in
this way, by iterating a process, will be called "canonical".
Suppose that, given the $k^{th}$ canonical partition, the nature
of the laws underlying the experiment is such that the
reiteration of the process will equidistribute values $x_i$ in
each and all segments in it; then we can continue defining $\mu$
in each of the new $2^{k+1}$ segments as $\frac{1}{2^{k+1}}$. As
$k \to \infty$, we obtain a probability measure on $[0,1]$,which
is, needless to say, singular with respect to the Lebesgue
measure.

\bigskip

Next, given a convenient value of $\al \ge 0$, we take a certain
$k$-partition whose segments will be called $I_i = I^{(k)}_i$,
will be enumerated from left to right and will have lengths $l_i =
l_i^{(k)}$. Now, we select those intervals $I_i$ for which
$\mu(I_i) \cong l_i^{\al}$; i.e. $\frac{\ln(\mu(I_i))}{\ln(l_i)}
\cong \al$, and we write $N_{\al} = N_{\al}^{(k)} = \# \{ i /
\frac{\ln(\mu(I_i))}{\ln(l_i)} \cong   \al \}$. When it is not
strictly necessary, we will drop the supra index $k$, writing $l_i$
or $I_i$ instead of $l_i^{(k)}...$ etc. Then we define $f(\al)$ as
the limit (if it exists), when $k \to \infty$ and the partition
gets finer and finer, of the quotient
$\frac{\ln(N_{\al})}{\ln(\frac{1}{l_i})}$. Notice that, if all
$\mu(I_i)$ in the same partition have the same value, then all
the $I_i$ selected so that $\frac{\ln(\mu(I_i))}{\ln(l_i)} \cong
\al$ have, albeit roughly, the same length ($l_i$).

\bigskip

This $(\al, f(\al))$ is obtained, when iterating the process
referred to above, in purely a computational way: we ask the
computer to select those $I_i$ such that
$\frac{\ln(\mu(I_i))}{\ln(l_i)} \cong \al$, we ask the computer to
count those indices $i \ldots$ We will, then, take the liberty of
calling $f_C$ ("$C$" for computational) this originally first
studied spectrum.

\subsection{The spectrum   $f_G(\al)$}

\medskip

While keeping all the structure just depicted for $f_C$, Riedi and
Mandelbrot [Riedi and Mandelbrot, 1998] stress that each and all partitions considered of
$[0,1]$ should be composed of segments of equal length. This is a
{\it most} sensible requirement, and it enables Riedi,
Mandelbrot and others to prove, in a rigorous way, different
properties about the corresponding spectrum, named $f_G$ ("$G$"
for grid).

Now, let us consider the measure described in Section 1.1,
equiprobable on each segment $I^{(k)}_i$ in canonical partition
$k$. Let us consider, in the same unit segment another partition,
a {\it uniform} one, composed of segments $I'_i$, all of
equal length. We can always fill up each segment $I'_i$ exactly,
with non-overlapping intervals $I^{(k)}$, with different values of
$k$, even if it is necessary to use very large values of $k$ in
order to do so. Then, we can obtain the exact value of $\mu(I'_i)\
\forall \ I'_i$ in the uniform partition. Once $\mu(I'_i)$ is known,
the rest of the structure proposed by Riedi and Mandelbrot for
$f_G$ is virtually the same as in 1.1 for $f_C$.

\bigskip

The (theoretical) distinction between the original $f_C$ and $f_G$ is
mathematically important, for $\frac{\ln(N_{\al})}{\ln(1/l_i)}$ has
no theoretically requirement to have a limit {\it independent
of the nature of the partition}.

\subsection{The inversion formula $\bar{f}(\al)$ of a spectrum $f(\al)$ }

\medskip
Let $(\al, f_H({\al}))$ be the Hausdorff spectrum of $(\Om, \mu)$, $f_H$ as 
defined in the Abstract.
Let us consider finer and finer canonical partitions $P_1,\ldots,
P_k,\ldots$. For each segment, $I_i^{(k)}$ we know its length and
measure $l_i^{(k)}$ and $\mu(I_i^{(k)})$ respectively. If we
wanted to invert the roles of $l_i$ and $\mu(I_i)$ |i.e. change
lengths with measures and vice versa| what would the new Hausdorff
spectrum, denoted by $(\al, \bar{f}_H(\al))$, look like? The
beautiful answer [Riedi and Mandelbrot, 1997], [Mandelbrot and Riedi, 1997], [Riedi and Mandelbrot, 1998] is the inversion formula
$\bar{f}_H(\al) = \al f_H(\frac{1}{\al})$.

The new segments in partition $k$ would have lengths
$\mu(I_i^{(k)})$ and measure $l_i^{(k)}$, so the new spectrum
should be very very different. Notice that the inversion formula
was proved by Riedi and Mandelbrot {\it for every probability measure!}

Riedi and Mandelbrot [Riedi and Mandelbrot, 1998] stress that if we replace $f_H$ by $f_G$, then
the inversion formula might not hold. Now, since $f_G= f_H$ [Riedi and Mandelbrot, 1998] for the
so called "self-similar measures", then the inversion formula holds,
trivially, for $f_G$ for such measures.

\subsection{Legendre coordinates and transformations}

\medskip
Let us recall that, when depicting $f_C$, we did not contemplate,
necessarily, a uniform partition. In [Halsey et al., 1986], however, the authors see
a very important advantage in working always with uniform
partitions: if $l = l^{(k)}$ is the length associated with the
intervals of a certain fine partition $P_k$, then the so-called
partition function $\sum \mu(I_i)^q$, $q$ a certain parameter
between $-\infty$ and $+\infty$, can be rewritten as $$\sum_i
\mu(I_i)^q = \sum_{\al} N_{\al}(l^{\al})^q = \sum_{\al}
N_{\al}l^{\al q}. $$ If we can confirm that $N_{\al} \sim
l^{-f_C(\al)}$ (which will certainly be the case for the Cantor
dust or any real self-similar fractal, for which $f_C = f_H$) we
would have for the partition function:

$$\sum_i \mu(I_i)^q = \sum_{\al}l^{-f_C(\al)} l^{\al q} =
 \sum_{\al}l^{(\al q -f_C(\al))}, $$ a quantity
very \ well \  estimated \ by $l^{\min_{\al}( \al q -
f_c(\al))}$,\ which  \ implies: $q-f_C'(\al)=0$, and for such
value of $q$ we would have $$\frac{\ln \sum p_i^q}{\ln l} \approx
\al q - f_C(\al),$$ where $p_i = \mu(I_i)$.

We call $\tau(q)$ the limit, when the norm of the partition $P_k$
tends to zero, of the logarithmic quotient above.

Needless to say, we terribly  oversimplified all the above
calculations |among other things it is not quite clear why $f_C$
would have a derivative.

The so-called Legendre transformations are then $\tau(q) \cong \al
q - f_C(\al)$, for a value of $q= f_C'(\al).$

Now, {\it if} the first of these two equations is really
tight enough |such as to replace "$\cong$" by "$=$"| then we can
differentiate both terms of it with respect to $q$: $$\tau'(q) =
\frac{d\al }{dq} q + \al - f_C'(\al).\frac{d \al}{dq} = \frac{d\al
}{dq} q + \al - q\frac{d \al}{dq} = \al$$ in our particular case
of $q = f_C'(\al)$.

So a third equation $\tau'(f_C'(\al)) = \al$ holds.

\subsection{Legendre's reading in Quantum Mechanics}

\medskip
Identifying $q$ with the inverse temperature $T$ of a system
(times the Boltzman constant $\beta$), $\al$ with its internal
energy $E$, $f_C(\al)$ with its entropy $S$, and relating the
partition function, via $\tau(q)$, with the free energy $F$ of the
system, then $$\tau(q) \cong \al q - f_C(\al) \  \mbox{and} \
f_C'(\al) =q $$ definitely becomes the classical relationships
between $T, E, S, F$, and the partition function [Duong-Van, 1987]; $\ldots$ so
that, theoretically, one could study some problems of Quantum
Mechanics as Fractal Geometry, and vice versa.

\bigskip

\begin{subsection}{The hyperbolic metric}

\medskip
We will very briefly touch on this subject, referring the interested
reader to standard texts. Let $\H$ be the open half plane $\{(x,y)
\in \R^2 / y >0\}$. In $\H$ we will define the rigid movements
(and they will be the equivalent of our Euclidean translations,
rotations,... all congruences) as follows: Let $(x,y)$ and
$(x',y')$ be in $\H$. If there is a $2 \times 2$ matrix $M$ with
real entries $a_{ij}$ and unit determinant such that
$\frac{y'}{x'} = \frac{a_{11}x + a_{12}y}{a_{21}x + a_{22}y}$,
then we say that $(x',y')$ is $(x,y)$ moved by a hyperbolic rigid
movement. The value of the trace $a_{11}+a_{22}$ determines the
nature of this movement. When all entries $a_{ij}$ are in $\zZ$
we define the set of these $2 \times 2$ matrices as the unimodular
(multiplicative) group $\bf{U}$ {\it which keeps appearing all over Mathematics when one least expects it!} To this $\bf{U}$ there is a tiling of $\H$
associated: there is a tile |called fundamental tile| $R$ such
that, if we apply each and all elements of $\bf{U}$ to $R$ we
obtain an infinity of non overlapping tiles (i.e. intersecting only at
their boundaries) covering all of $\H$.

There is a unique measure on $\H$ such that {\it all} these
infinite tiles have the same measure. Without entering into
detail let us call it $\mu^{\H}$, the hyperbolic measure. Each
tile is a perfect polygon; all tiles are, if we look at them with
hyperbolic spectacles, the same polygon, displaced, by rigid
movements, from place to place.

Let us suppose that we have two tiles $T_1$ and $T_2$, and that
each of them has two vertices $v^1_1, v^1_2$ and $v^2_1, v^2_2$,
respectively, in $[0,1]$. The unit segment inherits, then, from
$\mu^{\H}$ a certain measure $\mu$: since $T_1$ and $T_2$ are the
same for $\mu^{\H}$, in shape, size, and measure,  then $\overline{v^1_1
v^1_2}$ and $\overline{v^2_1 v^2_2}$ have to be, for the inherited
measure $\mu$, totally indistinguishable. {\it Par ab\'us de langage}
let us denote this new measure on the unit segment to be $\mu^{\H}$.

Now: the tiling associated with $\bf{U}$ is far from
 unique: Series
[Series, 1985] performed a surgical operation on the fundamental tile
$R$, obtaining another fundamental $\bar{R}$ and another tiling
(with interesting properties) associated with $\bf{U}$, also
covering exactly $\H$ with no overlapping, and inheriting another
hyperbolic measure on the unit segment. Elsewhere [Grynberg and Piacquadio, 1995] we have
operated on $\bar{R}$ producing another $\bar{\bar{R}}$ and
another tiling of $\H$ associated with $\bf{U}$, whose inheritance
$\mu^{\H}$ on the unit segment is, precisely, the Farey-Brocot
(F-B) $P_k$ partitions (see below). The segments in $[0,1]$ with
the same $\mu^{\H}$ measure (such as $\overline{v^1_1v^1_2}$ and
$\overline{v^2_1v^1_2}$ above) are now segments in the same
$P_k$ partition. The larger the $k$, the smaller the segments.

\bigskip

We are interested in the F-B partition of the unit segment,
since it crops up in a variety of problems in Physics.

\subsection{The Farey-Brocot partition in Physics}

\medskip
We propose to give here a variety of examples in which the F-B
partition appears. But first, let us illustrate the canonical
formations of the F-B partitions with a simple diagram:

\bigskip
Diagram A:
\bigskip

\bigskip

\setlength{\unitlength}{1cm}
\begin{picture}(12,5)
\put (0,5){$P_0 \hspace{.5cm}\frac{0}{1}$}

\put (1,5){\line(1,0){5.5}} \put(6.5,5){$\frac{1}{1}$}

\put(7,5) {$\mu^{\H}([0,1]) = 1$}

\put (0,3.1){$P_1 \hspace{.5cm} \frac{0}{1}$} \put
(1,3){\line(1,0){5.5}} \put(3.4,3.1){$\frac{0+1}{1+1}$} \put
(6.5,3.1){$\frac{1}{1}$}

\put (7,3){$\mu^{\H}([0,\frac{1}{2}]) = \mu^{\H}([\frac{1}{2},1]) =
\frac{1}{2^1}$}

\put (0,1.1){$P_2 \hspace{.5cm} \frac{0}{1}$} \put
(1,1){\line(1,0){5.5}} \put (3.5,1.1){$\frac{1}{2}$} \put
(2.4,1.1){$\frac{0+1}{1+2}$} \put(4.4,1.1){$\frac{1+1}{1+2}$}
\put(6.4,1.1){$\frac{1}{1}$}

\put(7,1){$\mu^{\H}([0, \frac{1}{3}]) = \mu^{\H}([\frac{1}{3},
\frac{1}{2}]) = $}

\put(7,0){$\mu^{\H}([\frac{1}{2}, \frac{2}{3}]) =
\mu^{\H}([\frac{2}{3}, 1]) =\frac{1}{2^2}$}

\put(0,0){$\vdots$}

\end{picture}

\bigskip

 We trust the diagram is self-evident, and that,
henceforth, we can replace $\mu^{\H}$ by a simplified notation $\mu$,
without confusion.

Let us now go to some physical examples.

\medskip

\subsubsection{Physical examples in which the F-B or hyperbolic
measure appears.}

\medskip

Let us develop three examples of physical phenomena where the
Farey-Brocot sequence (the Farey tree in the Physics literature)
appears, in the hope that these examples will convince the reader
of the relevance of the F-B sequence in Physics and Chemistry.

\bigskip

\item{ 1)} Let us consider the forced pendulum, with internal
frequency $\omega$. The angle $\theta$ formed by pendulum and
vertical is expressed by $\theta_{n+1} = f(\theta_n, \omega)$.
When plotting the winding number $W$ of \  $\theta_n$ as a
function $g$  of $\omega$, we have that, for certain critical
values of the parameters involved, $W= g(\omega)$ is a Cantor-like
staircase [Halsey et al., 1986]. It means that $g(\omega)$ is constant in the so
called intervals of resonance $I_k$ ($k$ a natural number) of the
variable $\omega$, each $I_k$ producing a step of the staircase.

Cvitanovic, Jensen, Kadanoff and Procaccia [Cvitanovic et al., 1985] discovered a
property of the staircase $W = g(\omega)$: Let $\frac{p}{q}$ and
$\frac{p'}{q'}$ be the values of $W$ for a pair of intervals of
resonance $I$ and $I'$, such that all intervals of resonance in
the gap between $I$ and $I'$ are smaller in size than both $I$ and
$I'$. Then, there is an interval of resonance $I"$ in this gap
such that the corresponding constant value of $W$ is
$\frac{p"}{q"} = \frac{p+p'}{q+q'}$. This interval $I"$ is the
widest of all intervals of resonance in the gap between $I$ and
$I'$. This is a purely empirical finding.

\bigskip

\item{ 2)} Bruinsma and Bak [Bruinsma and Bak, 1983] studied the one-dimensional Ising
model with long range anti ferromagnetic interaction |of strength
given by an exponent $\al >1$ | in an applied magnetic field $H$.
For zero temperature, the plot of $y = f_B(x)$, where $x = -H$, and
$y$ is the ratio of up spins over the total number of spins, is a
Cantor-like staircase. The steps of the staircase are the
stability intervals $\Delta H$ for which $y$ is a rational
number $p/q$. Again, these stability intervals [Piacquadio and Grynberg, 1998] are
distributed following the Farey tree described above for the
forced-pendulum Cantor-like staircase.

\bigskip
\noindent 3) To study the general properties of fermion systems
such as crystal formation, Falicov and Kimball proposed a
simple model. Gruber, Ueltschi and Jedrzejwki [Gruber et al., 1994] have
considered the one-dimensional Falicov-Kimball model. Let us
suppose that we have $q$ sites marked on the real line, $p'$ ions
to put in the $q$ sites and $p$ electrons hovering around the
ions.

Once the sites where the ions lie are chosen, we have a
configuration $\omega$ that we repeat periodically in order to
obtain an infinite one-dimensional lattice. The electron density
is $P_e = p/q$. Given $U$, the electron-ion interaction, and the
electrochemical potentials $(\mu_e, \mu_i)$ there is a
configuration $\bar{\omega}$ which minimizes the free energy
density $f(\omega, \mu_e,\mu_i)$. For small values of $U$, the
phase plane $(\mu_e, \mu_i)$ is divided into connected regions
$D_{P_e}$, such regions share the same $P_e$, and this $P_e$ is
the electronic density of an $\omega$ that minimizes $f(\omega,
\mu_e, \mu_i)$.

Numerical results show that between regions $D_{p/q}$ and
$D_{p'/q'}$ there appears a smaller one corresponding to $P_e =
\frac{p+p'}{q+q'}$. This region is the largest between $D_{p/q}$
and $D_{p'/q'}$.

\bigskip

Cvitanovic et al. [Cvitanovic et al., 1985], Halsey et al. [Halsey et al., 1986], and Rosen [Rosen, 1998] have
different examples of physical phenomena exhibiting Cantor
staircases with such (F-B) arrangements. We can also find this
arrangement in some of the staircases shown in [Bak, 1986], including the
chemical reaction of Belusov-Zabotinsky.
Other examples can be found in [Arrowsmith et al., 1996] and [McGehee and Peckham, 1996].

\subsection{The purpose of this paper}

\bigskip
\subsubsection{Self-similar measures and the hyperbolic measure}

\medskip
Reviewing a diversity of results, Riedi and Mandelbrot [Riedi and Mandelbrot, 1998] stress
the following:

\noindent{a)} For the so-called "self-similar measures" all {\it reasonably well defined} spectra
$f_{\al}$ coincide, and

\noindent{b)} fulfill the Legendre equations.

\noindent{c)} In the general case of other measures, $f_G$ was the
largest, $f_H$ the smallest function.

\noindent{d)} $f = f_H$ fulfilled the inversion formula for every
measure in $[0,1]$: $\bar{f}(\al) = \al f(1/\al)$, where $\bar{f}$
means, as we saw, that probability measures become lengths of
intervals and vice versa.

\noindent{e)} $f=f_G$ may not, in the general case of an arbitrary
measure, fulfill the inversion formula |and interesting examples
are given.

\medskip

The following diagram provides an example of  what a "self-similar
measure" is. Here, $a$ and $p$ are irrational numbers in $(0, 1)$.

\bigskip
Diagram B:
\bigskip

\bigskip

\setlength{\unitlength}{1cm}
\begin{picture}(10,5)
\put (0,4.5){$0$}\put (6.1,4.5){$1$}
\put(0,4.5){$\line(1,0){6.1}$} \put
(0,3){$\underbrace{\line(1,0){2}}_{a}$}
\put(2,3){$\underbrace{\line(1,0){4.2}}_{(1-a)}$} \put
(0,1.5){$\underbrace{\line(1,0){.5}}_{a^2}$} \put
(0.6,1.5){$\underbrace{\line(1,0){1.5}}_{a(1-a)}$}
\put(2.1,1.5){$\underbrace{\line(1,0){1.5}}_{(1-a)a}$}
\put(3.7,1.5){$\underbrace{\line(1,0){2.5}}_{(1-a)^2}$}

\put(6.5,4.5){$\mu([0,1]) = 1$}

\put(6.5,3){$\mu([0,a]) = p; \ \mu([a,1]) = 1-p$}

\put(6.5,1.6) {the measures of the corresponding}

\put(6.5,1.2){ intervals are, respectively,}

\put(6.5,0.8){$p^2, p(1-p), (1-p)p, (1-p)^2$}

\put(3,0){$\vdots$} 

%\put (3.5,0){and so on}

%\centerline{and so on}

\end{picture}

\centerline{and so on.}

\bigskip

\bigskip

Notice that the structure $\{1\}$; $\{a, 1-a \}$; $\{a^2, a(1-a),
(1-a)a, (1-a)^2\}\ldots$ etc. depicting the lengths of intervals
$I_i$, is formally {\it  identical} with that of $\{1\}$; $\{p, 1-p \}$;
$\{p^2, p(1-p), (1-p)p, (1-p)^2\}\ldots$ etc., depicting the
probability measures of said intervals $I_i$.

That is, when we replace the role of lengths of $I_i$ by its
probability measures $\mu_i$ we do not alter the formal structure:
we end up with

\bigskip
DIAGRAM C:

\setlength{\unitlength}{1cm}
\begin{picture}(13,5)

\put (2.1,4.5){$\bar{\mu}([0,p])=a$}

\put (1,4){$\underbrace{\line(1,0){3.7}}_{p}$}

\put (7,4.5){$\bar{\mu}([p,1])=1-a$}

\put (4.7,4) { $\underbrace{\line(1,0){6}}_{(1-p)}$}

\put (1,1.){$\underbrace{\line(1,0){1}}_{p^2}$}

\put (2,1) { $\underbrace{\line(1,0){2.5}}_{p(1-p)}$}

\put (4.7,1) {$\underbrace{\line(1,0){2.5}}_{(1-p)p}$}

\put (7.3,1) {$\underbrace{\line(1,0){3.5}}_{(1-p)^2}$}

\put(1,1.5){$\bar{\mu}=a^2$}

\put (2.5,1.5){$\bar{\mu}=a(1-a)$}

\put (5,1.5){$\bar{\mu}=(1-a)a$}

\put (8.2,1.5){$\bar{\mu}=(1-a)^2$}
\end{picture}
$\ldots$ etc.

\bigskip

\bigskip

Now, when we have an F-B arrangement of $[0,1]$ with the
hyperbolic measure, and we exchange roles for lengths and
measures, then the new structure is very very different.

\bigskip
For "self-similar measures" the case $l(I_i^{(k)}) =
(1-a)^{k-j}a^j$; for values of $j$ such that $0 \le j \le k$;
$\mu(I_i^{(k)}) = (1-p)^{k-j}p^j$ implied
$$\frac{\ln(l(I_i))}{\ln(\mu(I_i))} = \frac{1}{c + d \frac{j}{k}}
+ \frac{1}{\frac{k}{j}e + b}\ ,$$ where $b =
\frac{\ln(p/(1-p))}{\ln(a/(1-a))}$, $c =
\frac{\ln(1-p)}{\ln(1-a)}$, $d = \frac{\ln(p/(1-p))}{\ln(1-a)}$
and $ e = \frac{\ln(1-p)}{\ln(a/(1-a))}$; so for the inversion $l
\longleftrightarrow
 \mu$ we can write the new $\bar{l}_i$ and their corresponding measures
$\bar{\mu}_i$ directly, by selecting the $(i, j)$ order of the
segment in the partition.

\bigskip

\medskip
For our F-B arrangement |in this case $(\hbox{F-B})_3$ only:

\bigskip
DIAGRAM D:

\setlength{\unitlength}{1cm}
\begin{picture}(10,1)
\put (0,0){$\underbrace{\line(1,0){4}}_{\mu=\frac{1}{2^3}}$}

\put(0,.3){0}

\put (4,0){$\underbrace{\line(1,0){2.5}}_{\mu=\frac{1}{2^3}}$}

\put(4,.3){$\frac{1}{4}$}

\put (6.5,0){$\underbrace{\line(1,0){2}}_{\mu=\frac{1}{2^3}}$}

\put (6.5,.3){$\frac{1}{3}$}

\put (8.5,0){$\underbrace{\line(1,0){1.5}}_{\mu=\frac{1}{2^3}}$}

\put (8.5,.3){$\frac{2}{5}$}

\put (10,.3){$\frac{1}{2}$}

\end{picture}
\bigskip

\bigskip

\bigskip

\noindent There is no such possible easy way of understanding
$\bar{\mu}$, once we reverse lengths and measures.

The reader is invited to develop $(\hbox{F-B})_k$ arrangements for large
values of $k$ and to check how irregular the hyperbolic measure is
when compared to the Euclidean one. In fact, Series and Sinai
[Series and Sinai, 1990], among others, devoted excellent articles in order to stress
the irreconcilable differences |both physical and mathematical|
between the hyperbolic and the ordinary Lebesgue measure. For
instance, looking  at $\H$, one would say that, as a half plane, it
has dimension 2. Nevertheless, a common plane can be tessellated in
a specific and finite number of ways if all the convex tiles are
to be the same in shape and size. Other tilings are equivalent to
the latter. But the infinity of  |not equivalent| corresponding
tilings of $\H$ allows us to show that $\H$ behaves, in certain
ways, as $\R^{\infty}$.

Series and Sinai were also able to construct on $\H$ a non-numerable set of non-equivalent Gibbs measures.

\bigskip

We propose to show that, despite their fundamental differences,
the hyperbolic measure in $[0,1]$  behaves very much like the
self-similar ones in the sense of Section 1.8.1; that is:

\item{a')} For the hyperbolic measure, it appears that $f_G$ and $f_H$
coincide. If, as in c), $f_G$ is the largest and $f_H$ is the
smallest, then "all reasonably well defined spectra coincide" for
the hyperbolic measure, and

\item{b')} this measure fulfills the Legendre equations.

\bigskip

Some proofs are analytical, others are numerical. Both  a') and b')
as well as other results shown in Section 6 show a powerful analogy
between Euclidean and hyperbolic measures.

\medskip

We trust to be able to give some conjecture(s), at the end of the
paper, that could explain such analogies.

\section{$f_C$ and $f_G$ for the hyperbolic measure and the inversion formula for $f_C$}

\subsection{$f_C$ and $f_G$ for the hyperbolic measure}

\medskip

Let us recall, once more, how the F-B sequences were constructed;
let us consider them between $1/2$ and $1$, due to the symmetry of
the denominators. Henceforth we will write $\mu^{\H} = \mu$ for short.

\bigskip

Diagram E:
\bigskip

\setlength{\unitlength}{1cm}
\begin{picture}(12,5)

\put (0,4){$k=0 \hspace{0.5cm} \line(1,0){3.5}$ }

\put(1.3,4.1){0}

\put(4.8,4.1){1}

\put(4.8,2.6){1}

\put(4.8,1.1){1}

\put (0,2.5){$k=1 \hspace{1.2cm} \frac{1}{2} \line(1,0){2.5}
\hspace{.5cm} \mu([\frac{1}{2},1]) = \frac{1}{2^1} $}

\put (3,1.3){$\frac{2}{3}$}
\put(0,1){$k=2 \hspace{1.2cm} \frac{1}{2}  \line(1,0){2.5} \hspace{.5cm}
\mu([\frac{1}{2},\frac{2}{3}]) = \mu([\frac{2}{3},1]
=\frac{1}{2^2}$ }

\end{picture}

Now, we want a {\it uniform} $k$-partition of $[\frac{1}{2} ,
1]$, and we want to know the measure $\mu^{\H} = \mu$ of
{\it each} such interval in the uniform partition. We will
take a relatively small $k$, $k=11$, and we will divide $[0,1]$ in
equal segments of length $\frac{1}{2^{11}}$.

We stress that $k$ is rather small, since $[0,1]$ would be divided
into $2^{11} = 2048$ equal segments, and, as we will see
throughout the last sections, we need partitions of well over $4$
million segments in order to obtain a certain accuracy in our
numerical results. We will take the segment $I=[\frac{m}{2^k},
\frac{m+1}{2^k}], m = 1265$, in this partition, and we will try to
calculate $\mu(I)$. For
different F-B $k$-partitions, we will locate our interval $I$.
Briefly, the situation is as follows:

\bigskip

Diagram F:

\bigskip

%DIAGRAM F

\setlength{\unitlength}{1cm}
\begin{picture}(12,12)

\put (0,1){$k=11 \hspace{0.5cm} \line(1,0){2.2}$}

\put (7,1){\line(1,0){3.2} $ \hspace{.1cm} \mu(I)>2\times
\frac{1}{10} + 2 \times \frac{1}{2^{11}}$ }

\put (1,0){$\vdots$}

\put(1.5,1.2){$a $ here $\frac{97}{157} \hspace{0.5cm}
\frac{76}{123}$}

\put(6.5,1.2){$\frac{89}{144} \hspace{0.5cm} \frac{123}{199}$ $b$
here }

\put (0,3){$k=10 \hspace{0.5cm} \line(1,0){8.6} \hspace{.1cm}
\mu(I)>2\times \frac{1}{10} $ }

\put(1.5,3.2){$a $ here $\frac{76}{123}$}

\put(8,3.2){$\frac{89}{144}$ $b$ here }

\put (0,5){$k=9 \hspace{0.5cm} \frac{21}{34} \line(1,0){8.3}
\hspace{.1cm} \frac{34}{55} $ }

\put(5.5,5.2){$\frac{55}{89} \hspace{1cm} b=\frac{m+1}{2^{11}}$ here}
\put(2.5,5.2){$a=\frac{m}{2^{11}}$ here}

\put(5.5,3){\line(0,1){2}}

\put(0,7){$\vdots$}

\put (0,9){$k=4 \hspace{2cm} \frac{3}{5} \hspace{.1cm}
\line(1,0){4.6} \hspace{.1cm} \frac{2}{3} $ }

\put(5.5,9.2){$\frac{5}{8}$} \put(3.7,9.2){I here}

\put (0,10.8){$k=3 \hspace{.5cm} \line(1,0){8.6} $ } \put(1.1,11)
{$\frac{1}{2}$}

\put(3.9,11) {$\frac{3}{5}$}

\put(4.5,11.2){I included here}

\put(7.2,11) {$\frac{2}{3}$}

\put(10,11){1}

\put(5.5,2.9){$\underbrace{\hspace{2.5cm}}_{\mu =
\frac{1}{2^{10}}}$}

\put(3,2.9){$\underbrace{\hspace{2.5cm}}_{\mu =
\frac{1}{2^{10}}}$}

\put(7,.9){$\underbrace{\hspace{1.2cm}}_{\mu = \frac{1}{2^{11}}}$}

\put(2.6,.9){$\underbrace{\hspace{1.2cm}}_{\mu =
\frac{1}{2^{11}}}$}

\end{picture}

\bigskip

\bigskip
Proceeding in this way, we locate $a$ and $b$ for increasing
values of $k$; and, as shown in the column at the right, we
estimate $\mu(I)$ from below; but with increasing accuracy.
Finally, in $k= 37$ (and not before) we find both $a= \frac{m}{2^{11}}$ and $b =
\frac{m+1}{2^{11}}$, and we are able to give an {\it exact}
answer to the question "how much is
$\mu([\frac{1265}{2^{11}},\frac{1266}{2^{11}}])$?" Now, the
reality is that, in order to obtain the value of
$\mu([\frac{m}{2^{11}},\frac{m+1}{2^{11}}])$, the computer has to
reach, at least, to the F-B 37-partition... and the fastest
machine at our University gets stuck for days and days at
$k=24$.

Even if, with the help of some super computer (in some far away
place) we could possibly bridge the gap between $k=23$ and $k=37$,
we still would have a uniform partition (of intervals, together
with their exact $\mu^{\H}$ measures) of barely 2000 intervals...
 when we need over 4 million at the very very least.

\ldots We are absolutely forced to work, not with the
mathematically correct uniform partition, but with the F-B
partitions as they come |and that means that we are forced to work
with $f_C$ in lieu of the mathematically well defined $f_G$.

\subsection{The inversion formula for $f_C$}

 Working directly with $f_C$ instead of the well defined $f_G$ has
an advantage: the inversion formula holds for $f_C$, a result whose proof
is rather short: We know that, if $k$ is large enough,
then

$$\bar{f}_C(.) \cong \frac{ \ln(\# \{ i /
\frac{\ln(p_i^{(k)})}{\ln(1/2^k)} \cong . \} )}{\ln(1/2^k)}, $$
where both signs "$\cong$" above tend to "$ = $" as $k \to
\infty$. Therefore,

$$\al\bar{f}_C(\frac{1}{\al}) \cong \al \frac{ \ln(\# \{ i /
\frac{\ln(p_i^{(k)})}{\ln(1/2^k)} \cong \frac{1}{\al} \})
}{\ln(1/2^k)} \cong $$

$$ \cong \frac{\ln(1/2^k)}{\ln(p_i^{(k)})} \frac{ \ln(\#\{ i /
\frac{\ln(1/2^k)}{\ln(p_i^{(k)})} \cong \al \}) }{\ln(1/2^k)} =
\frac{\ln(\# \{ i / \frac{\ln(1/2^k)}{\ln(p_i^{(k)})} \cong \al \}
) }{\ln(p_i^{(k)})} = f_C(\al), $$ and, with "$\cong$" turning to
"$=$" as $k \to \infty$ we obtain $\al \bar{f}_C(\frac{1}{\al}) =
f_C(\al)$, or $\bar{f}_C(\frac{1}{\al}) = \frac{1}{\al} f_C(\al),
$ i.e., $\bar{f}_C(\al) = \al f_C(\frac{1}{\al})$, which is the
{\it inversion formula}.

\section{The theoretical spectrum $(\al, f_H(\al))$ for the
hyperbolic measure}

\medskip
Let us recall the definition of the theoretical spectrum $(\al,
f_H(\al))$ given in Section 1.

Let $x \in [0,1]$ and $I^{(k)}(x)$ be the unique interval in the
$k^{th}$ step of the F-B partition where $x$ lies; then

\begin{equation}
\label{uno}
\al(x) = \lim_{k \to \infty} \al_k(x) = \lim_{k \to
\infty} \frac{\ln(\mu(I^{(k)}{(x)}))}{\ln(l(I^{(k)}(x)))}
\end{equation}
when the limit exists. Here $l(I^{(k)}(x))$ is the length of the
interval $I^{(k)}(x)$, and as defined before, $\mu(I^{(k)}(x))
=\frac{1}{2^k}$, for $\mu$ is the hyperbolic measure.

\bigskip

Given any value $\al \in \R^+$, let $\Om_{\al}$ be the set of $x
\in [0,1]$ which share the $\al$-index, $\Om_{\al} = \{ x / \al(x)
= \al \}$. By definition $f_H(\al) = d_H(\Om_{\al})$, where
$d_H(\Om_{\al})$ denotes the Hausdorff dimension of $\Om_{\al}$.

To estimate such a curve $(\al, f_H(\al))$ we have to obtain an
adequate formula for the length of the intervals $I^{(k)}$.
Inevitably, we must introduce some formalism about continued
fractions.

\bigskip

Any real number $x \in (0,1)$ can be expressed as a continued
fraction:

$$
x ={1\over\displaystyle n_1+{1\over \displaystyle n_2+{1\over\displaystyle n_{3_{~~\displaystyle\ddots}}}}}
 = [n_1,
n_2, \ldots ], \  \mbox{with} \ n_i \in \N.$$

The sequence is finite if and only if $x$ is rational.

If $x$ is irrational, and we consider the $N^{th}$ rational
approximant to $x$:

$${P_{_N}\over Q_{_N}}={1\over\displaystyle n_1+{1\over\displaystyle n_2+_{~~\displaystyle\ddots_{\displaystyle {1\over\displaystyle n_{_N}}}}}}=
[n_1,n_2,...,n_{_N}],$$
 then $Q_N$ is the so-called "cumulant", a polynomial
in the variables $n_1, \cdots, n_N$.

If $x=[n_1, \ldots, n_N, \ldots]$, the length of $I^{(k)}(x)$ in
the $(\hbox{F-B})_k$ partition is $(Q_N.Q_{N-1})^{-1}$, provided
$k=n_1+n_2+\ldots +n_N$ [Cesaratto and Piacquadio, 1998].

Thus we have to estimate $Q_N$ in order to obtain
$\al(x)$. Obviously, $Q_N$ depends on the partial quotients $n_1,
\ldots, n_N$. The following lemma (demonstrated in [Cesaratto, 1999]) gives us
an adequate estimate of $Q_N$ as a function of $n_1, \ldots, n_N.$

\begin{Lemma}
Let $C_N$ be the set of rational numbers $[n_1, \ldots, n_N]$ for
which $n_i$ does not exceed $K$. Let us partition $C_N$ into
disjoint classes $C_N^{l_1 \ldots l_K}$, where $l_1, \ldots, l_K$
are in $\N$, $0 \le l_i \le N$, $l_1 + \ldots + l_K = N$. We will
say that $[n_1, \ldots , n_N] \in C_N^{l_1 \ldots l_K}$ if $l_1$
elements $n_i$ are equal to $1$, $l_2$ elements $n_i$ are equal to
$2$, \ldots etc. Let us define the frequency $\lambda_i =
\frac{l_i}{N}$, and rename the $N$-classes $C_N^{l_1 \ldots l_K}$ as
$C_N^{\lambda_1 \ldots \lambda_K}$.

If $[n_1, \ldots, n_N] \in C_N^{\lambda_1 \ldots \lambda_K}$, then
$\root N \of {Q_N(n_1, \ldots, n_N)}$ is well approximated by
$\sqrt{\pi^2/6 - 1}\ .2^{\lambda_1}\ldots (K+1)^{\lambda_K}$.
\end{Lemma}

\subsection{A connection between $\al(x)$, $x=[n_1, \ldots,n_N \ldots]$ and
$\lim_{N\to \infty} \frac{\sum^N_{i=1}n_i}{N}$}

Returning to the $\al$-formula (\ref{uno}) we see that, with $k,
\ K,$  and $N$ as before, $\al(x)$ can be rewritten as

\begin{equation}
\label{unoprima}
\begin{array}{lcl}
 \al(x) = \lim_{k \to \infty} \frac{k
\ln(2)}{\ln(Q_N Q_{N-1})} = \lim_{N \to \infty} \frac{\ln(2)}{2}
\frac{\sum^N_{i=1} n_i}{\ln(Q_N)} =   \\
 = \lim_{N \to \infty}
\frac{\ln(2)}{2} \frac{\sum^N_{i=1} n_i}{N} \frac{1}{
\ln(\sqrt{\pi^2/6 - 1}) + \sum^K_{i=1} \lambda_i \ln(i+1)}
\end{array}
\end{equation}

From (\ref{unoprima}) we notice that $\al(x), \ x=[n_1, \ldots, n_N,
\ldots]$ depends on the average $\frac{\sum^N_{i=1}n_i}{N}$ of the
partial quotients $n_i$. Therefore, we need to estimate the
Hausdorff dimension $d_H(F_m)$, where $F_m = \{ x \in [0,1] /
\lim_{N\to \infty} \frac{\sum^N_{i=1}n_i}{N} = m \}.$

We obtain [Cesaratto and Piacquadio, 1998], [Cesaratto, 1999]:

\item{ a)} $d_H(F_m) \cong d_H(F_m \cap E_{K_m}) =  1-{6\over \pi^2K_m}-
{72\ln(K_m)\over \pi^4K_m^2}$, where
$E_K = \{ x \in [0,1] / 1 \le n_i \le K \}$ , and $K_m \cong
e^{(\frac{\pi^2}{6} - 1)m}.$

\item{ b)} $d_H(F_m)$ is the Hausdorff dimension of the set of elements $x$
which belong to the $N$-classes $C_N^{\lambda_1 \ldots
\lambda_{K_m}}$ for which $$\lambda_i = \bar{\lambda_i} = \frac{
\frac{1}{(i+1)^{2d_H(E_{K_m})}} } {\sum^{K_m}_{i=1}
\frac{1}{(i+1)^{2d_H(E_{K_m})}}}.$$ Such a set, contained in $F_m$,
is large enough so as to be responsible for $F_m$ in Hausdorff
dimension and Hausdorff measure. Any $\lambda_i$-distribution
other than $\bar{\lambda_i}$ will correspond to a class of
Hausdorff dimension inferior to $d_H(F_m)$ |hence, of Hausdorff
measure nil in such a dimension.

Let us go back to (\ref{unoprima}): $$ \al(x) = \frac{\ln(2)}{2}
\lim_{N \to \infty }\frac{\sum^N_{i=1}n_i}{N}\frac{1}{\ln(\sqrt{
\pi^2/6 - 1}) + \sum^{K_m}_{i=1} \lambda_i \ln(i+1)} = m \bf{C},
$$ for short. The average of the $n_i$ is denoted by $m$, and ${\bf C} =
{\bf C} (\lambda_1  \ldots  \lambda_{K_m},  m)$.

Let us consider  $\lambda_i = \bar{\lambda}_i$. Then, the
stability of $\bf{C}$ depends on the stability of $\sum^{K_m}_{i=1}
\bar{\lambda}_i \ln(i+1) = (\sum_{i=1}^{K_m} \frac{1} { (1+i)^{ 2d_H(E_{K_m}) } }
)^{-1} \sum_{i=1}^{K_m} \frac{\ln(i+1)} {(i+1)^{2d_H(E_{K_m})}}$.
Notice that, if $m$ is not too small, then both sums
$\sum^{K_m}_{i=1}$ above are very much like $\sum^{\infty}_{i=1}
\frac{1}{(i+1)^2}$ and $\sum^{\infty}_{i=1} \frac{\ln(i+1)}{(i+1)^2}$
respectively. Therefore, if $m$ is not too small, the coefficient
${\bf C}(m)$ would be, essentially, independent of $m$, and we can
call it $\bf{C}$, for short. We would, then, obtain a connection
|via $\bf{C}$| between $\al(x)$ and $m$, when $x=[n_1, \ldots n_N,
\ldots ]$ and $\lim_{N \to \infty} \frac{\sum^N_{i=1} n_i}{N} =
m$.

\bigskip

Let us consider other elements $x = [n_1, n_2, \ldots ]$ in $F_m$.
For instance, $x_1 = [m, m, \ldots, m, \ldots]$ is, no doubt, in
$F_m$. But the corresponding ${\bf C}(x_1)$ is
$\frac{1}{\ln(\sqrt{\pi^2/6 - 1}) + \ln(m+1)}$, a number as small
as we want it, if $m$ is large enough.

Again, let us consider $x_2 = [1, 2m-1, 1, 2m-1, \ldots]$. We have
$\lim_{N \to \infty }\frac{\sum^N_{i=1}n_i}{N}= m,$ but
${\bf C}(x_2) = \frac{1}{\ln(\sqrt{\pi^2/6 - 1}) + \ln(2)
+\frac{\ln(m)} {2}}$, a number quite different from ${\bf C}(x_1)$,
if $m$ is large.

Again, we can construct another $x_3 = [n_1, n_2, \cdots ] \in
F_m$ with strings $(m, \cdots, m)$ and strings $(1, 2m-1, 1, 2m-1,
\ldots)$ large enough as to ensure that ${\bf C}(x_3)$ varies |with $N$|
from ${\bf C}(x_1)$ to ${\bf C}(x_2)$ and back an infinity of times $\cdots$.
Fortunately, all such elements  $x \in F_m$ that destabilize the
coefficient ${\bf C}(x)$ |moreover, such that possess a coefficient
different from ${\bf C}$| form a set of Hausdorff measure zero in the
Hausdorff dimension $d_H(F_m)$. Hence, with $\al = m {\bf C}$ we will
say that $\Omega_{\al}$ is, essentially, the set of elements $x$
in $F_m \cap E_{K_m}$ with distribution $\{
\bar{\lambda}_i\}_{i=1}^{K_m}$, and that $f_H(\al) =
d_H(\Omega_{\al}) = d_H(F_m) = d_H(F_m \cap E_{K_m}) =  1-{6\over \pi^2e^{({\pi^2\over 6} -1)m}}-
{72({\pi^2\over 6} -1)m\over \pi^4e^{2({\pi^2\over 6} -1)m}}$.

\subsection{Small values of $m=\frac{\sum^N_{i=1} n_i}{ N}$}

\medskip

{\bf Prior observations:}

\item{ 1)} Let us consider variables $m$ and $K_m$. $K_m$ is a
discrete variable, whereas $m$ is a continuous one. If both $K_m$
and $m$ are large enough, the difference between $d_H(F_m \cap
E_{K_m})$ and $d_H(F_m)$ will not show. What really happens is
that there are different values of the continuous variable $m$ |a
whole interval of such $m$| associated with the same $K_m$.

If $m$ and $K_m$ are large enough, these $m$ intervals are very
small$\ldots$ but for small values of $m$, say, for $m \in [1,
1.39]$ we have $K_m =1$! We should take into account that these
small values of $m$ are of great importance to physicists |even
if mathematicians tend to consider asymptotic behaviour| since,
for $m \in [1, 1.39]$ the corresponding $d_H$ varies from 0 to
0.677, i.e. more than half of the whole range of $f(\al)$!

\item{2) }$\lim_{K \to \infty} d_H(F_m \cap E_K)$ coincides with $d_H(F_m
\cap E_{K_m})$ for $m \ge 5$, but again the former is very different from the
latter for small values of $m$. For such small values of $m$, we obtain
a better estimate of $d_H(F_m)$ with $\lim_{K \to \infty} d_H(F_m
\cap E_K)$.

\item{ 3)} Finally, let us notice, in the last section, how many times we
had to write "$\ldots$ if $m$ is not small $\ldots$" or "$\ldots$
if $m$ is large enough $\ldots$"

All of which makes necessary a separate treatment for the case in
which $m$ is small.

The following theorem (we will not prove it here) gives us an estimate of $d_H(F_m)$ via
$\lim_{K \to \infty}d_H(F_m \cap E_K)$, which allows us to
treat the case $m >1$, $m$ small.

\bigskip

\begin{Theorem} Let $m$ and $F_m$ be as before. Let $K \in\N, \ K \ge 2$,
let $(y_{K,m} , z_{K,m})$ be the solution in $(e^{-2},1) \times
[0,1]$ of the following system of equations:
\begin{equation}
\label{system} \left\{
\begin{array}{lcl}
f^m_K(y,z) = \sum^{K-1}_{i=0} (m-i-1)y^{a_i}z^i = 0 \\
\\
g^m_K(y,z) =  \frac{y^t}{z^{m-1}} \sum^{K-1}_{i=0} y^{a_i}z^i - 1
= 0 \\
\end{array}
\right.
\end{equation}
where $a_i = (i-1)\ln(2) - i\ln(3) + \ln(i+2)$ and $t =
\ln(\sqrt{\pi^2/6 - 1}) - (m-2)\ln(2) + (m-1)\ln(3)$.

Let $\bar{y}_m = \lim_{K \to \infty} y_{_{K, m}}$. Then $\lim_{K \to
\infty}d_H(F_m \cap E_K) = -1/2 \ln(\bar{y}_m)$.

\end{Theorem}

\bigskip

Moreover, let $K_m$ be large enough as to ensure $\bar{y}_m \cong
y_{K_m,m}$. Let us write $\bar{\bar{\lambda}}_{i,K} = y_{K,m}^{a_i
+t}z_{K,m}^{i+1-m}$. Then, $\lim_{K \to \infty} d_H(F_m \cap E_K)
= -\frac{1}{2}\ln(\bar{y}_m)$ agrees with the dimension of $
C^{\bar{\bar{\lambda}}_{1, K_m}, \cdots,
\bar{\bar{\lambda}}_{K_m,K_m}},$ which we abbreviate as
$C^{\bar{\bar{\lambda}}_{1}, \cdots, \bar{\bar{\lambda}}_{K_m}} =
\{ x = [n_1, \cdots n_N, \cdots] / \lim_{N \to \infty} \frac{l_i}{N}
= \bar{\bar{\lambda}}_{i,K_m} \}$.

\bigskip

For all $x \in C_N^{\bar{\bar{\lambda_1}}, \ldots,
\bar{\bar{\lambda_{K_m}}}}$ we have $$\al = \al(x) =
\frac{\ln(2)}{2} \frac{ \sum_i i\bar{ \bar{ \lambda_i} } } { \ln(
\sqrt{\pi^2/6 -1 }) + \sum_{i=1}^{K_m} \bar{\bar{\lambda}_i} \ln(
i+1)}=$$
$$=\frac{\ln(2)}{2} \frac{m}{ \ln(
\sqrt{\pi^2/6 -1 }) + \sum_{i=1}^{K_m} \bar{\bar{\lambda}_i} \ln(
i+1)}\cong\al(m).$$

For this value of $\al = \al(m)$ we have that $f_H(\al) =
d_H(\Omega_{\al}) = d_H(F_m) = -1/2 \ln(\bar{y}_m)$. In Table 1 we
show some numerical values obtained as we have just explained.

\pagebreak

%Tabla 1

\begin{center}
TABLE 1: Some numerical values for the theoretical spectrum.

\smallskip

\begin{tabular}{|c|c|c|c|c|c|}
\hline
  m    & K   & $y_{K,m}$ & $z_{K,m}$ & $\alpha$ & $f_H(\alpha)$ \\ \hline
  1.01 & 10  & 0.8892    & 0.0099    & 0.7325   & 0.0587        \\ \hline
  1.02 & 20  & 0.8152    & 0.0196    & 0.7336   & 0.1021        \\ \hline
  1.04 & 20  & 0.7074    & 0.0383    & 0.7358   & 0.1730        \\ \hline
  1.08 & 20  & 0.5689    & 0.0733    & 0.7404   & 0.2821        \\ \hline
  1.2  & 20  & 0.3744    & 0.1596    & 0.7561   & 0.4912        \\ \hline
  1.4  & 25  & 0.2582    & 0.2561    & 0.7863   & 0.6770        \\ \hline
  1.8  & 30  & 0.1863    & 0.3531    & 0.8562   & 0.8402        \\ \hline
  2    & 30  & 0.1717    & 0.3780    & 0.8942   & 0.8809        \\ \hline
  2.4  & 50  & 0.1560    & 0.4070    & 0.9736   & 0.9290        \\ \hline
  3    & 60  & 0.1456    & 0.4268    & 1.0988   & 0.9633        \\ \hline
  3.5  & 70  & 0.1416    & 0.4346    & 1.2061   & 0.9773        \\ \hline
  4    & 100 & 0.1393    & 0.4386    & 1.3173   & 0.9854        \\ \hline
  5    & 300 & 0.1371    & 0.4419    & 1.5528   & 0.9936        \\ \hline
  6    & 300 & 0.1362    & 0.4437    & 1.6879   & 0.9968        \\ \hline
  7    & 400 & 0.1358    & 0.4443    & 2.032    & 0.9983        \\   \hline
\end{tabular}

\medskip

\end{center}

The spectrum $(\al, f_H(\al))$ has the graph shown in Fig. 1:

\medskip

\begin{center}

%\begin{figure}
\includegraphics[width=8cm,height=5cm]{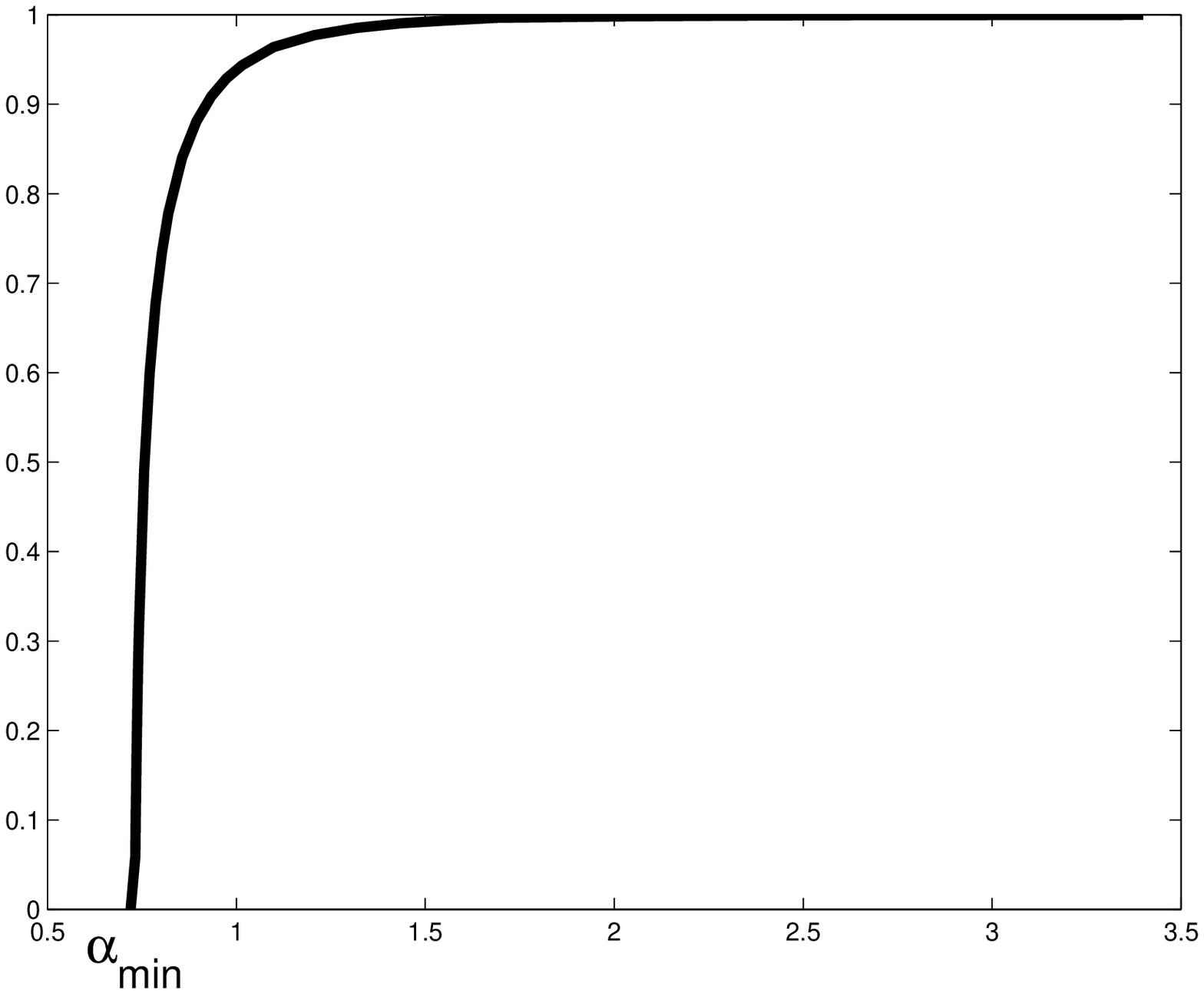}

%\caption{The graph of $f_H(\al)$. Notice that $\al_{min} =
%\frac{\ln(2)}{\ln(\varphi^2)}; \  \phi = \frac{\sqrt5 - 1}{2}$}
%\end{figure}
Fig. 1: The graph of $f_H(\al)$. Notice that $\al_{min} =
\frac{\ln(2)}{\ln(\varphi^2)}; \ \varphi = \frac{\sqrt5 - 1}{2}$ [Grynberg and Piacquadio, 1995].

\smallskip
\end{center}

\section{The computational spectrum $( \al, \\ f_C (\al) )$}

From now on $k$ will enumerate the F-B sequences, and $p_i =
p_i^{(k)}$ will denote $l(I_i^{(k)})$.

We have $f_C(\al) = \lim_{k \to \infty}  f_C^{(k)}(\al) $ where $
f_C^{(k)}(\al) = \al \bar{f}_C^{(k)} (\frac{1}{\al})$;
$\bar{f}_C^{(k)}$ the computational function obtained, via
Legendre conditions, from  $\bar{\tau}^{(k)}(q) =
\frac{\ln(\sum^{2k}_{i=1} p_i^q)}{\ln(\frac{1}{2^k})}$, where $p_i
= p_i^{(k)} = \frac{1}{Q_i^{(k)}Q_{i+1}^{(k)}}$ and $I_i^{(k)} = [
\frac{P_i^{(k)} }{Q_i^{(k)}},
\frac{P_{i+1}^{(k)}}{Q_{i+1}^{(k)}}]$.

Now, when plotting  $f_C^{(k)}(\al)$ versus $\al$ we notice that,
$\forall k$, the spectrum is a perfect type III of Tel [Tel, 1988], as seen in
Fig. 2.

\bigskip

%\begin{figure}
\begin{center}
\includegraphics[width=10cm,height=6.5cm]{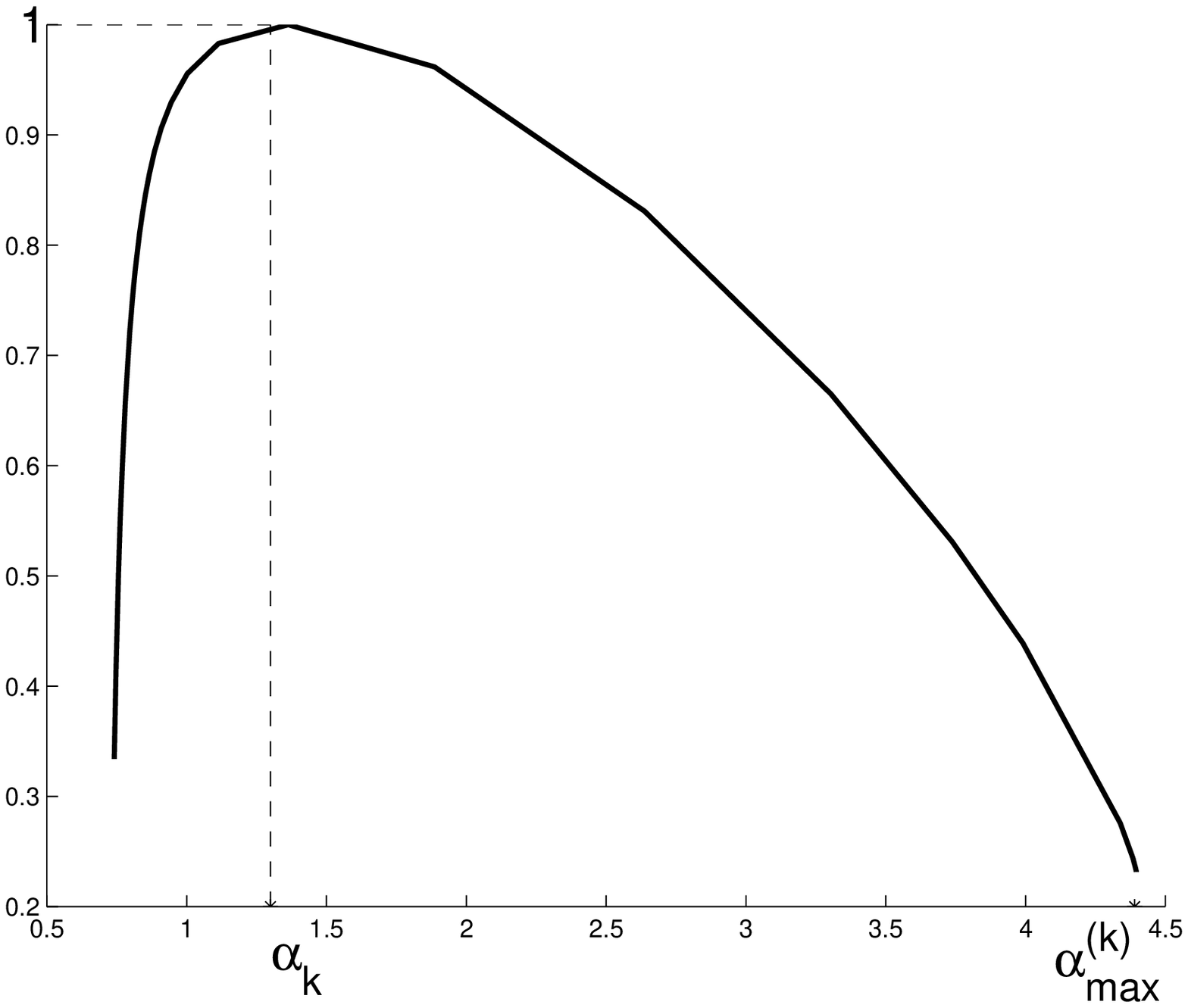}
%\caption{}

Fig. 2: The spectrum $f_C^{(k)}(\al). $
%\end{figure}
\end{center}

\bigskip

$f_C^{(k)}(\al)$ always reaches the value unity at $\al = \al_k$,
and then decreases to a minimun reached at $\al^{(k)}_{max}$. The
distance $\Delta^{(k)}\al = \al^{(k+1)}_{max} -
\al^{(k)}_{max}$ decreases to zero as $k$ increases, so we are
tempted to conclude that $f_C = \lim_{k \to \infty} f_C^{(k)}$ is
also type III. Yet, $\Delta^{(k)}{\al}$ goes to zero much slower
than the harmonic $\frac{1}{k}$, as shown in Table 2:

\medskip

\begin{center}

TABLE 2: Some numerical values for the $\Delta^{(k)}\al$.

$\Delta^{(k)}\al \gg \frac{1}{k};$
$\Delta^{(k)}\al\simeq \frac{1}{\sqrt{k+25}}.$

\smallskip

%tabla 2
\begin{tabular}{|c|c|c|}
\hline
  $k$ & $\al_{max}$ & $\Delta^{(k)}\al$ \\ \hline
  10  & 2.709       & 0.1814               \\ \hline
  11  & 2.8906      & 0.1777               \\ \hline
  12  & 3.0684      & 0.1745               \\ \hline
  13  & 3.2403      & 0.1716               \\ \hline
  14  & 3.4144      & 0.1690               \\ \hline
  15  & 3.5834      & 0.1666               \\ \hline
  16  & 3.75        & 0.1644               \\ \hline
  17  & 3.9144      & 0.1624               \\ \hline
  18  & 4.0768      & 0.1605               \\ \hline
  19  & 4.2374      & 0.1588               \\ \hline
  20  & 4.3962      & 0.1572               \\ \hline
  21  & 4.5534      & 0.1557               \\ \hline
  22  & 4.7091      &                      \\ \hline
\end{tabular}

\end{center}

\medskip

So, in fact, $\al^{(k)}_{max} \to +\infty$.

\medskip
Let us examine the

\subsection{Theoretical reasons responsible for $\al^{(k)}_{max}
\to +\infty$}

\medskip
Let us first prove 

\begin{Theorem}
\label{teorema2}
No matter how large $\al$ is, $f_C(\al) > 0.$
\end{Theorem}

\noindent{\bf Sketch of the proof.}

We know that $f_C(\al) = \lim_{k\to \infty} \frac{ \ln(\#\{ i \ /
\ \frac{ \ln(\frac{ 1}{2^k})}{\ln(p_i^{(k)})} \cong \al \}) }{
\ln( \frac{ 1}{ p_i^{(k)}}) } = $ $$= - \lim_{k \to \infty}\frac{
\ln( \# \{ i \ / \  \ln(p_i^{(k)}) \cong \frac{ \ln(\frac{
1}{2^k})}{\al} \})}{ \frac{\ln( \frac{ 1}{ 2^k})}{\al})} =$$ $$=
\al \lim_{k \to \infty} \frac{ \ln(\# \{ i \ / \
\ln(Q_i^{(k)}Q_{i+1}^{(k)}) \cong \frac{k\ln(2)}{\al} \})}{k
\ln(2)} =$$ $$= \frac{\al}{\ln(2)} \lim_{k\to \infty} \frac{\ln(
\# \{ i\ / \ \frac{\ln(Q_i^{(k)}) + \ln(Q_{i+1}^{(k)}) }{k} \cong
\frac{\ln(2)}{\al} \})}{k},$$ so $f_C(\al)$ away from zero implies
$ \frac{\ln( \# \{ i\ / \ \frac{\ln(Q_i^{(k)}) +
\ln(Q_{i+1}^{(k)}) }{k} \cong \frac{\ln(2)}{\al} \})}{k}:=
{\bf Q}^{(\al)}(k)$ away from zero if $k \ge k(\al)$ is large enough.

Let us notice that, if $\al$ is very large, such quotient will be
very small. Let us call it $\ve(\al)$. Therefore, the numerator of
${\bf Q}^{(\al)}(k)$ will behave as $\ve(\al)k$ for $k$ large enough. Then
$\# \{ \ \}$ behaves as $e^{\ve(\al)k}$, which in turn implies
$$\frac{\# \{ i / \frac{\ln(Q_i^{(k+1)}) +
\ln(Q_{i+1}^{(k+1)})}{k+1} \cong \frac{\ln(2)}{\al} \} }{\# \{ i /
\frac{\ln(Q_i^{(k)}) + \ln(Q_{i+1}^{(k)})} {k} \cong
\frac{\ln(2)}{\al} \} }  \approx \frac{e^{\ve(\al)(k+1)}}
{e^{\ve(\al)k}} = e^{\ve(\al)} > 1,$$ $e^{\ve(\al)}$ being
independent of $k$.

But, if $\al$ is very large, even if we take large values of $k$,
it could be that $\ve(\al)$ is so small that $e^{\ve(\al)}$,
though greater than unity, will be so close to unity as to appear
indistinguishable form it. Let us analyse the reasons behind the
extreme smallness of $\ve(\al)$ when $\al$ is large. The condition
$$ \frac{\ln(Q_i^{(k)}) + \ln(Q_{i+1}^{(k)})}{k} \cong
\frac{\ln(2)}{\al}$$ can be rewritten as $$
\frac{\ln(Q_i^{(k)})}{k} + \frac{\ln(Q_{i+1}^{(k)})}{k} \cong
\frac{\bar{\pi}}{M},$$ where $M$ is a suitable constant, and
$\bar{\pi}$ is the limit, as $k \to \infty$, of the average of the
$2^k$ logarithms of the bases $Q^{(k)}(i)$ of the F-B
denominators $Q_i^{(k)}$ considered as exponentials with exponent
$k,$ i.e.  $Q^{(k)}_i = [Q^{(k)}(i)]^k$. Such average
$$\frac{\sum_{i=1}^{2^k}\ln(Q^{(k)}(i))}{2^k} = \bar{\pi}_k$$
increases with $k$, $\bar{\pi}_k \to \bar{\pi} \ (k \to \infty),$
a value that cannot exceed $\ln(\varphi)$, since $\varphi^k$ is a very
good approximation of the largest F-B denominator in the
$k^{th}$ step. The constant $M = \frac{\bar{\pi}}{\ln(2)}\al$ is a
measure of how large $\al$ is. Now, one of the two values,
$Q_i^{(k)}, Q_{i+1}^{(k)}$ is the largest of the pair, say
$Q_{i+1}^{(k)}$. Therefore, $Q^{(k)}(i)$ varies growing from
unity, in which case its logarithm is zero and
$\frac{\ln(Q_{i+1}^{(k)})}{k} \cong \frac{\bar{\pi}}{M}$, reaching
a value of $Q_{i}^{(k)}$ very near $Q_{i+1}^{(k)}$, in which case
$\frac{\ln({Q_{i+1}^{(k)}})}{k}\cong \frac{\bar{\pi}}{2M}$. In both
cases, and in general, $\frac{\ln({Q_{i+1}^{(k)}})}{k}\cong \lambda
\frac{\bar{\pi}}{M}, \ \lambda \in (\frac{1}{2}, 1]$ (and $\frac{
\ln{(Q_i^{(k)}})}{k}$ is even smaller). If we have in mind that the
average of such values $\frac{\ln{(Q_i^{(k)})}}{k}$ is $\bar{\pi}$,
then we understand that, to talk about $$"\frac{\ln{(Q_i^{(k)})}}{k}
= \lambda \frac{\bar{\pi}}{M} "$$is to talk about a minute
quantity of indices $i$, since $\lambda \frac{\bar{\pi}}{M}$ is a
tiny proportion of $\bar{\pi}$.

Intuitively, it may be clear that such quantity of ''$i$" will grow
with $k$, but such growth could be so slow as to be undetectable
from $k$ to $k+1$.

We will introduce two simplifications, one of them quite coarse:
we will be quite generous with the sign "$\cong$" in
$$\frac{\ln(Q_i^{(k)}) + \ln(Q_{i+1}^{(k)})}{k} \cong
\frac{\bar{\pi}}{M}:$$ this will mean for us
$$\frac{\ln(Q_i^{(k)}) + \ln(Q_{i+1}^{(k)})}{k} \in
[\frac{\bar{\pi}}{M}(1- \frac{1}{4}) , \frac{\bar{\pi}}{M}(1
+\frac{1}{4})].$$

Even now, if $\al$ (and therefore $M$) is very very large, the
corresponding quantity of indices $i$ verifying such condition is
so minute that we cannot detect its growth from $k$ to $k+1$.
Therefore, we will look for an $m = m(\al)$ so large such that,
from $k$ to $k+ m(\al)$ we will detect such growth of the quantity
of "$i$"; i.e. the quotient between $\# \{ i /
\frac{\ln(Q_i^{(k+m)}) + \ln(Q_{i+1}^{(k+m)})}{k+m} \cong
\frac{\bar{\pi}}{M} \}\ \hbox{ and  } \# \{ i / \frac{\ln(Q_i^{(k)}) +
\ln(Q_{i+1}^{(k)})}{k} \cong \frac{\bar{\pi}}{M} \}$ will be
clearly seen as larger than unity.

In order to achieve that,  it will be enough to take
{\it one} F-B interval with denominators $Q_i^{(k)}$ and
$Q_{i+1}^{(k)}$ |which we will denote, for simplicity, $I_{k,i} =
[Q_i^{(k)}, Q_{i+1}^{(k)}]$ | which fulfills the condition
$\frac{\ln(Q_i^{(k)}) + \ln(Q_{i+1}^{(k)})}{k} :={\bf q}_{i,k} \in [
\frac{\bar{\pi}}{M}(1-\frac{1}{4}),\frac{\bar{\pi}}{M}(1+\frac{1}{4})]
:=Int,$ and verify that, for a certain $m = m(\al)$ sufficiently
large we have, from the step $k+ m(\al)$ onwards, {\it two}
intervals $I_{k+ m(\al), j}$ (in $I_{k,i}$) for which the
corresponding ${\bf q}_{j, k+m(\al)}$ belong to $Int$.

We start from ${\bf q}_{i,k} \in Int$, for a pair $(Q_i^{(k)},
Q_{i+1}^{(k)}) $ of denominators adjacent in F-B, say, of the
order of $e^{\frac{3}{4} \frac{\bar{\pi}}{2M}k}$ and
$e^{\frac{5}{4} \frac{\bar{\pi}}{2M}k}$ respectively: the
corresponding ${\bf q}_{i,k}$ is $\frac{3}{8} \frac{\bar{\pi}}{M} +
\frac{5}{8} \frac{\bar{\pi}}{M} = \frac{\bar{\pi}}{M}$ exactly.

We want to see if, for steps $k+1, k+2, \ldots, k+h,\ldots$ it is
easy to find an interval $I_{k+h,j}$ inside $I_{k,i}$ that
inherits ${\bf q}_{i,k+h} \in Int$.

Let us F-B interpolate in $[Q_i^{(k)}, Q_{i+1}^{(k)}]$: we
obtain two intervals $L$ and $R$ (for left and right). From here
on $L^2$ will be the interval "left of the left", $LR$ "right of
the left", \ldots and so on.

We start by considering $L = [e^{\frac{3}{4}\frac{\bar{\pi}}{2M}k}
; e^{\frac{3}{4} \frac{\bar{\pi}}{2M}k} + e^{\frac{5}{4}
\frac{\bar{\pi}} {2M}k}];$ the interpolating value
$e^{\frac{3}{4}\frac{\bar{\pi}}{2M}k} + e^{\frac{5}{4}
\frac{\bar{\pi}}{2M}k}$ can be rewritten as $e^{\frac{5}{4}
\frac{\bar{\pi}}{2M}k}(1 + \frac{1}{e^{\frac{\bar{\pi}}{4M}k}}).$

We choose now $k= k(\al)$, our starting value of $k$: we want it
to be of the order of
$\frac{4M}{\bar{\pi}}\ln(\frac{4M}{5\bar{\pi}})$ (a huge number),
so that ${\frac{1}{e^{\frac{\bar{\pi}k}{4M}}}}$ is $\frac{5\bar{\pi}}{4M} $,
a very small number. Now $L =
[e^{\frac{3}{4}\frac{\bar{\pi}}{2M}k};
e^{\frac{5}{4}\frac{\bar{\pi}}{2M}k}(1+ \frac{5\bar{\pi}}{4M})]$
and the corresponding {\boldmath $q$} (with a slight change of notation that,
we hope, simplifies the text) is

$${\bf q}_{L,k+1} = \frac{\frac{3}{4}\frac{\bar{\pi}}{2M}k + \frac{5}{4}
\frac{\bar{\pi}}{2M}k + \ln(1+ \frac{5\bar{\pi}}{4M})}{k+1} =
\frac{\bar{\pi}}{M} \frac{k}{k+1} + \frac{\ln(1+
\frac{5\bar{\pi}}{4M})}{k+1}= $$ $$=\frac{\bar{\pi}}{M}
\frac{1}{k+1} [k +\ln(1+ \frac{5\bar{\pi}}{4M})
\frac{M}{\bar{\pi}}] \cong \frac{\bar{\pi}}{M} \frac{1}{k+1}
[k+\frac{5}{4}] =$$ $$= \frac{\bar{\pi}}{M}(1 + \frac{1/4}{k+1}) =
\frac{\bar{\pi}}{M} + \frac{\bar{\pi}}{4M(k+1)},$$ where
$\frac{\bar{\pi}}{4M(k+1)}$ is so very small that ${\bf q}_{L,k+1}$ is
almost indistinguishable from $\frac{\bar{\pi}}{M} : \ {\bf q}_{L,k+1}
\in Int.$

Let us explore the situation in $L^2 =
[e^{\frac{3}{4}\frac{\bar{\pi}}{2M}k};
e^{\frac{3}{4}\frac{\bar{\pi}}{2M}k} +
e^{\frac{5}{4}\frac{\bar{\pi}}{2M}k}( 1 +
\frac{5\bar{\pi}}{4M})].$ The interpolating value can be rewritten
as $e^{\frac{5}{4}\frac{\bar{\pi}}{2M}k }( 1 + 2
\frac{5\bar{\pi}}{4M}).$

Therefore, ${\bf q}_{L^2, k+2} = \frac{1}{k+2}[\frac{3}{4}
\frac{\bar{\pi}}{2M} k + \frac{5}{4} \frac{\bar{\pi}}{2M} k +
\ln(1+2 \frac{5 \bar{\pi}}{4M})] \cong (\frac{\bar{\pi}}{M} k + 2
\frac{5 \bar{\pi}}{4M})\frac{1}{k+2} = \frac{\bar{\pi}}{M}
\frac{1}{k+2} (k + \frac{5}{2}) = \frac{\bar{\pi}}{M}
\frac{1}{k+2} (k + 2 + \frac{1}{2}) = \frac{\bar{\pi}}{M} +
\frac{\bar{\pi}} {2M(k+2)}$, again a value very much alike
$\frac{\bar{\pi}}{M}$, slightly larger than it.

Next, we need to know if these ${\bf q}$ are increasing, getting away
from $\frac{\bar{\pi}}{M}$ towards $\frac{\bar{\pi}}{M}(1
+\frac{1}{4}),$ i.e. is $\frac{\bar{\pi}}{4M(k+1)} <
\frac{\bar{\pi}}{2M(k+2)}$? The answer is "yes". So, "${\bf q} \in Int$"
for $L, L^2, \ldots$ but up to a point, since the corresponding
${\bf q}$'s appear to increase |very slowly| and, at some distant future
they could surpass $\frac{\bar{\pi}}{M}(1+\frac{1}{4})$: let us
explore this situation $L^i$ and the corresponding ${\bf q}_{L^i,k+i}$.
$L^i$ is $[e^{\frac{3}{4} \frac{\bar{\pi}}{2M}k },  e^{\frac{5}{4}
\frac{\bar{\pi}}{2M}k }(1 + i\frac{5\bar{\pi}}{4M})]$, and
${\bf q}_{L^i,k+i} = \frac{ \frac{3}{4} \frac{ \bar{\pi}}{2M}k  +
\frac{5}{4} \frac{\bar{\pi}}{2M}k  + \ln(1+i
\frac{5\bar{\pi}}{4M})}{k+i} = \frac{\bar{\pi}}{M} \frac{k +
\frac{M}{\bar{\pi}} \ln(1 + i\frac{5\bar{\pi}}{4M})}{k+i}, \ldots$
and applying L'Hopital in the variable $i$ indicates that this
proportion of $\frac{\bar{\pi}}{M}$ eventually tends to zero! Now,
before the value ${\bf q}_{L^i, k+i}$ descends {\it below}
$\frac{\bar{\pi}}{M}(1 - \frac{1}{4})$ let us stop the process: we
choose $i$ such that ${\bf q}_{L^i, k+i}$ is of the order
$\frac{\bar{\pi}}{M}\frac{3}{4}$. We have $i = i(k) = i(k(\al))$.
From here we can conclude $\frac{ \frac{M}{\bar{\pi}} \ln(1 + i
\frac{5\bar{\pi}}{4M})}{k+i} = \frac{3}{4} - \frac{k}{k+i}$, and
$\frac{M}{\bar{\pi}} \ln(1 + i \frac{5\bar{\pi}}{4M}) =
\frac{3}{4}(k+i) - k$, and we can replace $i$ by $i+1$ (in this
equality) with confidence.

Then ${\bf q}_{L^iR, k+i+1} = \frac{\bar{\pi}/M}{k+i+1} [ \frac{5}{4}
\frac{1}{2}k + \frac{M}{\bar{\pi}} \ln(1+ (i+1) \frac{5\bar{\pi}}
{4M})] +   \frac{\bar{\pi}/M}{k+i+1} [ \frac{5}{4} \frac{1}{2}k +
\frac{M}{\bar{\pi}} \ln(1+ i \frac{5\bar{\pi}} {4M})] = \frac{
\bar{\pi}}{4M} \frac{1}{k+i+1} [5k + 4 (\frac{3}{4}(k+ i + 1)- k)
+ 4 (\frac{3}{4} (k+i) - k)] > \frac{3}{4} \frac{\bar{\pi}}{M} =
{\bf q}_{L^i, k+i}.$

For $L^iR, L^iR^2, \ldots, L^iR^h, \ldots$ the corresponding ${\bf q}$'s
start to grow again, with the danger of surpassing, eventually,
$(1+ \frac{1}{4})\frac{\bar{\pi}}{M}$.

However, ${\bf q}_{L^iR^h, k+i+h} = \frac{\frac{5}{4}
\frac{\bar{\pi}}{M} k + \ln( 1+ i\frac{5\bar{\pi}}{4M} ) + \ln( h
+(hi + 1)\frac{5\bar{\pi}}{4M} )}{k+i+h} = $ \\
$=\frac{\bar{\pi}}{M} \frac{\frac{5}{4}k + \frac{M}{\bar{\pi}}\ln(
1+i\frac{5\bar{\pi}}{4M}) + \frac{M}{\bar{\pi}}\ln(h
+(hi+1)\frac{5\bar{\pi}}{4M} )}{ k + i +h },$ and again, L'Hopital
in the variable $h$ shows that, eventually, the last quotient
diminishes until it reaches $1 - \frac{1}{4} = \frac{3}{4},
\ldots$ By choosing carefully the spelling, the letters are $L$
and $R$, of the word interval inside $L$ (notice that the spelling
is far from unique), we can reasonably be assured of finding an
interval inside $L$ for which the corresponding ${\bf q} \in Int$,
provided the step $k \ge k(\al)$.

\bigskip

\noindent{\bf The interval $R$}

\medskip
We will explore next the other segment inside $I_{k,i}$, i.e. let
us explore $R$ and the corresponding ${\bf q}_{R, k+1}$.

We have $R = [e^{\frac{5}{4} \frac{\bar{\pi}}{2M}k} (1 + \frac{
5\bar{\pi}}{4M}) ,  e^{\frac{5}{4} \frac{\bar{\pi}}{2M}k}]$, and
${\bf q}_{R, k+1} =  \frac{1}{k+1} [ \frac{5}{4}\frac{\bar{\pi}}{2M}k +
\ln(1+\frac{5\bar{\pi}}{4M}) + \frac{5}{4} \frac{\bar{\pi}}{2M}k]=
 \frac{1}{k+1} [ \frac{5}{4} \frac{\bar{\pi}}{M}k +
\ln(1+\frac{5\bar{\pi}}{4M})] < \frac{1}{k+1} [\frac{5}{4}
\frac{\bar{\pi}}{M}k + \frac{5\bar{\pi}}{4M}]$ (since $\ln(1+\ve)
< \ve) = \frac{5}{4} \frac{\bar{\pi}}{M}(k + 1) \frac{1}{k+1} =
\frac{5\bar{\pi}}{4M} $ exactly. Therefore $R$ does inherit (via
$\ln(1+\ve) < \ve$) the property ${\bf q} \in Int$.

Nevertheless, if we had approximated $\ln(1+\ve)$ by $ \ve$ (and
$\ve = \frac{5\bar{\pi}}{4M}$  is {\it very} small), we would
have had ${\bf q}_{R,k+1} = (1 + \frac{1}{4}) \frac{\bar{\pi}}{M}$, and
if we had considered $Int$ as open, then ${\bf q}_{R,k+1}$ would have
been simply {\it out} of $Int$. So $R$ would be out, would
not inherit ${\bf q} \in Int$ from $I_{k,i}$.

Let us consider next $RL$  and $R^2$. The corresponding F-B
interpolating element separating both segments is $e^{\frac{5}{4}
\frac{\bar{\pi}}{2M}k} (2 + \frac{ 5\bar{\pi}}{4M})$, and looking
at the extremes of both segments $RL$ and $R^2$, clearly the
smallest possible corresponding ${\bf q}$ will be the one associated
with $R^2$: ${\bf q}_{R^2, k+2}= \frac{\frac{5}{4} \frac{\bar{\pi}}{M}k
+ \ln(2 + \frac{ 5\bar{\pi}}{4M})}{k+2}$, larger still than
${\bf q}_{R,k+1} \cong \frac{5}{4} \frac{\bar{\pi}}{M}$.

\noindent $\ldots$And yet, considering $R^i$ we would have
${\bf q}_{R^i, k+i} = \frac{\bar{\pi}}{M} \frac{\frac{5}{4} k +
\frac{M}{\bar{\pi}}{ \ln(i + \frac{5}{4}
\frac{\bar{\pi}}{M})}}{k+i}, $ and, L'Hopital in variable $i$
shows that the last quotient can be made as small as we want.
Therefore, there is a value of $i$ such that ${\bf q}_{R^i, k+i} \in
Int$, say, ${\bf q}_{R^i, k+i} = \frac{\bar{\pi}}{M}$; and this $i$ is
our $m(\al)$. Then we change the spelling, by introducing letters
$L\ \ldots$and so on: we feel that, for $k \ge k(\al) + m(\al)$ we
can safely be assured of finding {\it two} intervals inside
the starting one $I_{k,i}$ for which the corresponding ${\bf q}$ is in
$Int\ \ldots$which is what we wanted.

\subsection{The $\al_k$ for which $f_C^{(k)} (\al_k) = 1$}

Let us go back to Fig. 2: $\al_k$ is the value of $\al$ for which
$f_C^{(k)}$ reaches unity. We {\it do} have that
$\bar{\Delta}^{(k)}\al = \al_{k+1} - \al_{k} \to 0$ when $k \to
+\infty$. Nevertheless $\bar{\Delta}^{(k)}\al \to 0$ far slower
than the harmonic $\frac{1}{k}$, as seen in Table 3.

\medskip

%tabla 3
\begin{center}

TABLE 3: Some numerical values for $\bar{\Delta}^{(k)}\al$.

$\bar{\Delta}^{(k)}\al \gg \frac{1}{k};$
$\bar{\Delta}^{(k)}\al \approx \frac{\ln(k+16)}{k+16}.$

\smallskip

\begin{tabular}{|c|c|c|}
\hline
  $k$ & $\al_k/\ f(\al_k)=1$  & $\bar{\Delta}^{(k)}\al$  \\ \hline
  10  & 1.2220   & 0.0178                      \\ \hline
  11  & 1.2398   & 0.0167                      \\ \hline
  12  & 1.2564   & 0.0157                      \\ \hline
  13  & 1.2722   & 0.0149                      \\ \hline
  14  & 1.2871   & 0.0141                      \\ \hline
  15  & 1.3012   & 0.0135                      \\ \hline
  16  & 1.3147   & 0.0129                      \\ \hline
  17  & 1.3276   & 0.0123                      \\ \hline
  18  & 1.3399   & 0.0118                      \\ \hline
  19  & 1.3518   & 0.0114                      \\ \hline
  20  & 1.3632   & 0.0110                      \\ \hline
  21  & 1.3741   & 0.0106                      \\ \hline
  22  & 1.3847   &                             \\ \hline
\end{tabular}

\end{center}

\medskip

Therefore $\al_k \to \infty$.

Let us explore the theoretical reasons underlying this fact.

\begin{Theorem} The value $\al_k$ for which $f_C^{(k)}$ is unity tends to
infinity as $k \to \infty$.
\end{Theorem}

\noindent {\bf Sketch of the proof}

 We know that $f_C^{(k)} (\al) = \al
\bar{f}_C^{(k)} ( \frac{1}{\al})$, where $\bar{f}_C^{(k)} $ is
obtained by changing weights $\frac{1}{2^k}$ by lengths $p_i^{(k)}
= \frac{1}{Q_i^{(k)}Q_{i+1}^{(k)}}$, having now a uniform
partition of the unit segment. For $\al = \al_k$ we have
$f_C^{(k)}(\al_k) = 1$ and $f_C^{(k)'}(\al_k) = 0$.

But
\begin{equation}
\label{dos} f_C^{(k)'}(\al_k) = \bar{f}_C^{(k)}(\frac{1}{\al}) -
\frac{1}{\al}\bar{f}_C^{(k)'}(\frac{1}{\al}) = 0
\end{equation}
for $\al = \al_k.$

Being obtained from a uniform partition, $\bar{f}_C^{(k)}$ is
expressible through the Legendre equations via a certain
$\bar{\tau}(q) = \bar{\tau}_k(q) = \frac{\ln(\sum^{2^k}_{i=1}
(p_i^{(k)})^q )}{\ln(\frac{1}{2^k})}$ (it makes no sense
complicating the notation by writing $\bar{q}$ or $\bar{q}_k$
instead of $q$). Therefore, with $\frac{1}{\al} = \beta$ we have

\begin{equation}
\left\{
\begin{array}{lcl}
\bar{\tau}_k(q) = \beta q - \bar{f}_C^{(k)}(\beta) \\
q=\bar{f}_C^{(k)'}(\beta) \\ \bar{\tau}_k'(q)=\beta
\end{array}
\right.
\end{equation}
which, from (\ref{dos}) and for $\beta_k = \frac{1}{\al_k}$
becomes

\begin{equation}
\left\{
\begin{array}{lcl}
\bar{\tau}_k(q) = 0                     \\
q=\bar{f}_C^{(k)'}(\beta_k)                 \\
\bar{\tau}_k'(q)=\beta_k
\end{array}
\right.
\label{beta}
\end{equation}

Now, $\bar{\tau}_k(q) = 0$ means $\sum^{2^k}_{i=1}(p_i^{(k)})^q =
1,$ which happens when $q = 1,$ and from (6) we obtain
$\bar{f}_C^{(k)'}(\beta_k) = 1$ and $\bar{\tau}_k'(1) = \beta_k.$

Let us consider $\bar{\tau}_k'(q) = \frac{\frac{1}{\sum_i
(p_i^{(k)})^q}  \sum_i(p_i^{(k)})^q  \ln(p_i^{(k)})}{
\ln(\frac{1}{2^k})}$; when $q=1$ we are left with

$$\bar{\tau}_k'(1) =  \frac{\sum_{i=1}^{2^k} p_i^{(k)}
 \ln(p_i^{(k)})}{-k \ln(2)} = \frac{
\sum_{i=1}^{2^k}\frac{\ln(Q_i^{(k)}
Q_{i+1}^{(k)})}{Q_i^{(k)}Q_{i+1}^{(k)}} }{k \ln(2)} = \beta_k.$$

Now, if it happens that $\beta_k \to 0$ when $k \to \infty$, then
$\al_k = \frac{1}{\beta_k} \to \infty$ when $k \to \infty$, which
is what we wanted.

Now, $\beta_k$ is of the order of $\frac{2}{\ln(2)}
\sum_{i=1}^{2^k}\frac{ \ln(Q_i^{(k)})}{kQ_i^{(k)}Q_{i+1}^{(k)}}$;
and we know that the average $\frac{
\sum_{i=1}^{2^k}\frac{\ln(Q_i^{(k)})}{k}}{2^k} = \bar{\pi}_k$
increases to $\bar{\pi} < \ln(\varphi) \cong 0.4812$.

Next, we notice that $\bar{\pi}_{18} = 0.3914, \bar{\pi}_{20} =
0.3917, \bar{\pi}_{22} = 0.3921, \ldots$ and if we study the type
of growth of $\bar{\pi}_k$ we safely conclude that $\bar{\pi}$ is
somewhere between $0.4$ and $0.4812$: $\bar{\pi}$, and therefore,
$\bar{\pi}_k$, when $k$ is very large, is surprisingly near its
maximal value $\ln(\varphi)$. Let us notice that
 this nearness of $\bar{\pi}_k$ to $\ln(\varphi)$ cannot be achieved on account
of compensating pairs of very unequal numbers such as, e.g.
$\frac{\ln( Q_i^{(k)})}{k} = \frac{\ln(1)}{k} = 0$ and
$\frac{\ln(Q_j^{(k)})}{k} = 2 \bar{\pi}_k$, since $2\bar{\pi}_k$ is
considerably larger than $\ln(\varphi)$, and said $Q_j^{(k)}$ is,
therefore, non existing. This observation, and a moment of
reflexion on the nature of the growth of the $Q_i^{(k)}$ when $k
\to \infty$, tell us that, if $k$ is very large, we may assume
that there is a certain percentage $p$ of $Q_i^{(k)}$ for which
$\frac{\ln(Q_i^{(k)})}{k} $ and $\bar{\pi}_k$ are quite close. The
tighter the closeness, the smaller the percentage $p$.

Then, if $k$ is large enough, save for a constant $c \in (p,1)$,
the order of $\sum^{2^k}_{i=1} \frac{\ln(Q_i^{(k)})/k}
{Q_i^{(k)}Q_{i+1}^{(k)}}$ is given by $2^k
\frac{\bar{\pi}_k}{e^{2k\bar{\pi}_k} },$ and
therefore the order of $\beta_k$ is given by $\frac{2}{\ln(2)}
\bar{\pi}_k \frac{2^k}{e^{2k\bar{\pi}_k}} = \frac{2
\bar{\pi}_k}{\ln(2)}[\frac{2}{e^{2\bar{\pi}_k}}]^k$.

$\bar{\pi}_k$ is stable for $k$ large. We are interested in what
happens to $\frac{2}{e^{2\bar{\pi}_k}}$ when $k$ is large, i.e. we
want to know if $2 < e^{2\bar{\pi}_k}$ or $\ln(2) < 2\bar{\pi}_k$
or $\frac{2\bar{\pi}_k}{\ln(2)} > 1$ if $k$ is large. When $k=22$,
we do obtain a value of $\bar{\pi}_k$ for which
$\frac{2\bar{\pi}_k}{\ln(2)} = 1.1314$, substantially larger than
unity$\ldots$ But notice that the corresponding order of the F-B
partition of the unit segment $[0,1]$ is over $4$ million
segments! We have, then, $\beta_k \to 0$ and $\al_k \to \infty$,
as we wanted.

\bigskip

Next, let us see that

\subsection{ $f_C''(\al) \le 0$ for $\al$ in its domain}

\medskip

We know that $f_C(\al) = \al \bar{f}_C(\frac{1}{\al})$, where
$\bar{f}_C = \lim_{k \to \infty} \bar{f}^{(k)}_C$, is a
spectrum with uniform partitions on the unit interval, and
therefore fulfills all the Legendre conditions, including
$\bar{f}_C''(\al) \le 0$ for any $\al$ in its domain.

Now $f_C'(\al) = \bar{f}_C(\frac{1}{\al}) + \al
\bar{f}_C'(\frac{1}{\al}) \frac{-1}{\al^2} =
\bar{f}_C(\frac{1}{\al}) - \bar{f}_C'(\frac{1}{\al}) \frac{1}{\al}
:=  \bar{f}_C(\beta) - \beta \bar{f}_C'(\beta)$; therefore
${f}_C''(\al) = \frac{d f_C'(\al)}{d\beta} \frac{d
\beta}{d\al} = [\bar{f}_C(\beta) - \beta
\bar{f}_C'(\beta)]'(-\frac{1}{\al^2}) = -\frac{1}{\al^2}
(\bar{f}_C'(\beta) - \bar{f}_C'(\beta) - \beta \bar{f}_C'' (\beta))
= \frac{1}{\al^2} \beta \bar{f}_C''(\beta) = \frac{1}{\al^3}
\bar{f}_C''(\beta) \le 0 $ since $\al \ge 0 $ always.

\subsection{What we know about $f_C(\al)$:}

\item{1)} It is zero in $\al = \frac{\ln(2)}{\ln(\varphi^2)}$ [Grynberg and Piacquadio, 1995].

\item{2)} For any $\al$ as large as we want it, it is strictly larger
than zero.

\item{3)} $f_C(\al) = \lim_{k\to\infty} f_C^{(k)}(\al)$, and  $f_C^{(k)}(\al_k) =
1$, with $\al_k \to \infty$ when $k \to \infty$.

\item{4)} $f_C''(\al) \le 0$ in its domain.

\bigskip

$1)$ to $4)$ indicate that, qualitatively, the shape of the curve
$(\al, f_C(\al))$ is the one for $(\al, f_H(\al))$: both curves
have, qualitatively, the same shape. Let us compare then now
computationally.

\section{The Legendre Equations and the hyperbolic measure}

\medskip
We know ([Cawley and Mauldin, 1992], [Riedi and Mandelbrot, 1998]) that, for a so called "self similar measure" we
have $f_C(\al) = f_H(\al)$, and both are equal to $\min_q(q\al -
\tau(q))$ |an expression that Mandelbrot defined as $f_L$.

Having a uniform partition in mind, $l$ being the length of each
little segment in the partition, and $\tau(q)$ being $\lim_{l \to
0} \frac{\ln(\sum_i p^q_i)}{\ln(l)}$, let us recall that the
first Legendre transformation is [Halsey et al., 1986]

\begin{equation}
\tau(q) = \min_{\al}(q\al - f_C(\al)) \label{tres}
\end{equation}
which implies $\frac{d}{d\al}(q\al - f_C(\al)) = 0$, i.e. $q =
f_C'(\al)$, and therefore $\tau(q) = q\al - f_C(\al)$ for $q =
f_C'(\al)$. Now, if we can differentiate both sides of this equality we
have $\tau'(q) = \al + q \frac{d\al}{dq} - f_C'(\al)
\frac{d\al}{dq} = \al$, since $q = f_C'(\al)$.

Then $f_C(\al) = q\al - \tau(q)$ for $q = f_C'(\al)$, and
$\tau'(q) = \al$ for the same value of $q$.

Let us now consider the other Legendre transformation: $f_L(\al) =
\min_q(q\al - \tau(q))$. This entails $\frac{d}{dq}(q\al -
\tau(q)) = 0$, i.e. $\al = \tau'(q)$. Therefore
\begin{equation}
\label{cuatro} f_L(\al) = q\al - \tau(q)
\end{equation}
where now $\al$ and $q$ are related through $\al = \tau'(q)$. If we
can differentiate equation (\ref{cuatro}) we obtain $f_L'(\al) =
\frac{dq}{d\al}\al + q - \tau'(q) \frac{dq}{d\al} =
\frac{dq}{d\al}\al + q - \al \frac{dq}{d\al} = q$, and therefore
we can write $f_L(\al) = q \al - \tau(q)$,  where $\al$ and $q$
are related through $f_L'(\al) = q$, or else $\tau(q) = q\al -
f_L(\al)$ for $q = f_L'(\al)$; which means that, if we can
differentiate equalities (\ref{tres}) and (\ref{cuatro}) |both relating
a spectrum $f(\al)$ with $\tau(q)$| then both Legendre
transformations are equivalent.

Now, the expression "differentiate an equality" has the following sense
here: originally [Halsey et al., 1986], equation (\ref{tres}) was a good approximation:
$\tau(q)$ was very well approximated by $q\al - f_C(\al)$ with $q
= f_C'(\al)$, and two functions can be very much alike \ldots not
so their derivatives. We will have |Section 6| opportunity to
stress the relevance of this point.

\subsection{The spectrum $(\al, f_H(\al))$ of F-B does fulfill
the Legendre transformation}

In order to refer to $f_L$ at all, we have to have a uniform
partition in mind, so we will work with $\bar{f_H}(\al) = \al
f_H(\frac{1}{\al})$, and verify that, for the corresponding
$\bar{\tau}(q) = \lim_{k \to \infty} \frac{\ln(\sum_{i=1}^{2^k}
p^q_i)}{\ln(1/2^k)}$ we have  $\bar{f}_H(\al) = \min_q (q\al -
\bar{\tau}(q))$ or, as we just saw,

\begin{equation}
\label{cinco}
\left\{
\begin{array}{lcl}
\bar{\tau}(\bar{f}_H'(\al)) = \bar{f}_H'(\al)\al - \bar{f}_H(\al)  \\
\bar{\tau}'(\bar{f}_H(\al)) = \al
\end{array}
\right.
\end{equation}

We prefer to write the Legendre transformation in the form
(\ref{cinco}) rather than (\ref{cuatro}) because of the following
reason: The RHS of (\ref{cinco}) is obtained through purely
theoretical means ([Cesaratto and Piacquadio, 1998], [Cesaratto, 1999] and Theorem (3.1)
%\ref{teorema1}
 above).
However, to know the value of $\tau(q)$ entails knowing the
expression of $p_i = p_i^{(k)}$ for every $i$ and $k$,  $p_i^{(k)}
= \frac{1}{Q^{(k)}_i Q^{(k)}_{i+1}}$; therefore we have to know
the value of $Q^{(k)}_i \ \forall i , \forall k$ |or, at least, a
good approximation to such a value. Since this is a yet unsolved
problem, the LHS of (\ref{cinco}) is obtained through purely
numerical methods: one finds the value of the $Q^{(k)}_i$ for the
largest $k$ that our computer can handle, and then one obtains the
best possible approximation to $\bar{\tau}(q)$ for chosen different
values of $q = \bar{f}_H'(\al) $ |which {\it are}
theoretically obtained.

\bigskip
For short, let us denote $\phi(\al) = f_H(\frac{1}{\al}),$ a
function we will deal with in Section 6. What we know about
$\phi$: the domain of $\phi$ is $[0, \frac{\ln(\varphi^2)}{\ln(2)}]$;
$\phi(0) = 1$; $\phi'$ is negative if $\al \ne 0$;
$\phi'(\frac{\ln(\varphi^2)}{\ln(2)})$ is a huge negative number $-M$
(possibly $-\infty$?); $\phi$ decreases from $1$ to zero as $\al$
goes from $0$ to $\bar{\al}_{max} = \frac{\ln(\varphi^2)}{\ln(2)}$;
$\phi'(0) = \phi''(0) = \ldots = \phi^{(h)}(0) = \ldots = 0$; $\phi$
is not analytic at $\al = 0$.

Now, $\bar{f}_H(\al) = \al \phi(\al)$; $\bar{f}_H'(\al) =
\phi(\al) + \al \phi'(\al)$.

\bigskip

We want to compare
\begin{equation}
\label{seis} \frac{\ln(\sum_{i=1}^{2^k} p_i^{\bar{f}_H'(\al)})}{-k
\ln(2)} \ \mbox{ \ with \ } \ \al\bar{f}_H'(\al) - \bar{f}_H(\al)
\end{equation}
for a high value of $k$. Recall that the left hand side of
(\ref{seis}) is numerically obtained, whereas the RHS is
theoretically obtained; $\bar{f}_H'(\al) = \phi(\al) +
\al\phi'(\al)$, and  $\al\bar{f}_H'(\al) - \bar{f}_H(\al) = \al
\phi(\al) + \al^2\phi'(\al) - \al \phi(\al) = \al^2 \phi'(\al)$.

With  $\bf{X}(\al)$ we denote the LHS and with $\bf{Y}(\al)$ the
RHS of (\ref{seis}), and we plot the curve
$(\bf{X}(\al), \bf{Y}(\al))$ for a diversity of values of $\al$.
The curve is indistinguishable from the line ${\bf Y }={\bf  X}$, and, in
order to make a more analytical comparison, we present Table
4:

\medskip

%Tabla 4

\begin{center}

TABLE 4 : Some numerical values for LHS and RHS of Eq. (10).

To calculate ${\bf Y}(\al_n)$ we considered $\bar{\al}_n
= \frac{\al_n + \al_{n+1}}{2}$ and $\bar{f}_H(\bar{\al}_n) =
\frac{\bar{f}_H(\al_n) + \bar{f}_H(\al_{n+1})}{2}, $ thereby
ending with 14 values of ${\bf X}(\al_n)$ and ${\bf Y}(\al_n)$.

\smallskip

\begin{tabular}{|c|c|c|c|c|c|c|}
\hline
  n & $\al_n$ & $\bar{f}_H(\al_n)$ & $\phi(\al_n)$ & $\phi'(\al_n)$ & LHS$={\bf X}(\al_n)$ & RHS$={\bf Y}(\al_n)$ \\ \hline
  1 & 0.4921  & 0.4913         & 0.9983        & -0.015         & -0.0079             & -0.044              \\ \hline
  2 & 0.5925  & 0.5906         & 0.9968        & -0.0621        & -0.0330             & -0.0237             \\ \hline
  3 & 0.6440  & 0.6339         & 0.9936        & -0.0712        & -0.0467             & -0.0348             \\ \hline
  4 & 0.7591  & 0.7480         & 0.9854        & -0.1157        & -0.0877             & -0.0728             \\ \hline
  5 & 0.8291  & 0.8103         & 0.9773        & -0.1729        & -0.1477             & -0.1305             \\ \hline
  6 & 0.9101  & 0.8767         & 0.9633        & -0.2931        & -0.2945             & -0.2740             \\ \hline
  7 & 1.0271  & 0.9542         & 0.9290        & -0.5274        & -0.6261             & -0.6058             \\ \hline
  8 & 1.183   & 0.9851         & 0.8809        & -0.82          & -1.0871             & -1.0711             \\ \hline
  9 & 1.1680  & 0.9813         & 0.8402        & -1.5718        & -2.3704             & -2.3348             \\ \hline
 10 & 1.2718  & 0.8610         & 0.6770        & -3.6577        & -6.1776             & -6.1523             \\ \hline
 11 & 1.3226  & 0.6496         & 0.4912        & -7.9559        & -13.3682            & -13.3185            \\ \hline
 12 & 1.3506  & 0.3810         & 0.2821        & -12.9209       & -23.8365            & -23.7174            \\ \hline
 13 & 1.3591  & 0.2351         & 0.1730        & -17.3957       & -32.4206            & -32.2273            \\ \hline
 14 & 1.3631  & 0.1392         & 0.1021        & -21.2014       & -39.7022            & -39.4545            \\ \hline
 15 & 1.3652  & 0.081          & 0.0587        &                &                      &                    \\ \hline

\end{tabular}

\end{center}

\medskip

The relative errors $\Delta_n = \frac{{\bf Y}(\al_n) -
{\bf X}(\al_n)} {\min\{{\bf Y}(\al_n), {\bf X}(\al_n)\}}$ estimated
by excess are shown in Table 5.

For values $n = 1, 2, 3, $ and $ 4$ the magnitudes of both
${\bf X}(\al_n)$ and ${\bf Y}(\al_n)$ are very small, and the
estimate of $\Delta_n$ is always misleading; i.e. let us suppose
that ${\bf X}(\al_n) = \ve$ and ${\bf Y}(\al_n) = \ve^3$. Hence
$\Delta_n$, estimated from below, is $\frac{\ve - \ve^3}{\ve} \cong 1$; and,
estimated by excess |as we did in our table| $\Delta_n $ is $\frac{\ve -
\ve^3}{\ve^3} \cong \frac{1}{\ve^2}$!

We will deal separately with the values of $n$ for which $\al_n$
is small |i.e. for which ${\bf X}(\al_n)$ and ${\bf Y}(\al_n)$ are
very small.

%tabla 5
\medskip

\begin{center}

TABLE 5: The errors $\Delta_n = \frac{{\bf Y}(\al_n) -
{\bf X}(\al_n)}{\min \{ {\bf Y}(\al_n),\ {\bf X}(\al_n) \}}.$

\smallskip

\begin{tabular}{|c|c|c|c|c|c|c|c|}
\hline
  $n$        & 1      & 2      & 3      &      4 & 5      & 6      & 7      \\
\hline
  $\Delta_n$ & 0.8190 & 0.3952 & 0.3420 & 0.2043 & 0.1320 & 0.0750 & 0.0335 \\ \hline
\end{tabular}

\bigskip

\begin{tabular}{|c|c|c|c|c|c|c|}
\hline
  8      & 9      & 10     & 11     & 12    & 13    & 14     \\
\hline
  0.0150 & 0.0153 & 0.0041 & 0.0037 & 0.005 & 0.006 & 0.0063 \\ \hline
\end{tabular}

\medskip

\end{center}

So we can safely conclude that equality (\ref{seis}) does follow.

\subsection{Theoretical aspects of equation (\ref{seis})}

Equality (\ref{seis}) can be rewritten as
\begin{equation}
\label{siete} \frac{\ln(\sum_{i=1}^{2^k} p_i^{\phi(\al) + \al
\phi'(\al)})}{-k \ln(2)} = \al^2 \phi'(\al)
\end{equation}
where the value of $k$ in the LHS is the largest one we can
handle numerically. Nevertheless, if the exponent $\phi(\al) + \al
\phi'(\al) = q$ takes certain key values, one would be able to see
why, analytically, equation (\ref{seis}) holds when $k \to
\infty$:

\item{1)} Let $\al =0$. Then $\phi(\al) = 1$ and $\phi'(\al) = 0$:
the corresponding exponent $q$ is unity, then $\sum^{2^k}_{i=1}
p_i^q = \sum^{2^k}_{i=1} p_i^{(k)} = 1$, and its log is zero,
hence $\bar{\tau}(q) = 0$. The RHS is $\al^2\phi'(\al)$, obviously
zero.

\item{2)} Let $\al = \frac{\ln(\varphi^2)}{\ln(2)} =
\bar{\al}_{max};$ $\phi(\bar{\al}_{max}) = 0;$
$\phi'(\bar{\al}_{max}) = -M,$ $M$ a huge number, the largest |in
absolute value| that $\phi'$ can have, i.e. that $\bar{f}_H'$ can
have. Now, let us recall that this largest possible negative value
of $q$ (i.e. of $\bar{f}_H'$) selects [Halsey et al., 1986] the smallest $p_i =
p_i^{(k)}$ in the partition function, which corresponds to
$p_i^{(k)} = \frac{1}{Q_i^{(k)}Q_{i+1}^{(k)}} \cong
\frac{1}{\varphi^{2k}}.$

The LHS of (\ref{siete}), therefore, is $$ \frac{
\ln([\frac{1}{\varphi^{2k}}]^{0+\frac{\ln(\varphi^2)}{\ln(2)}(-M)}) }{
-k \ln(2)} = \frac{-M \frac{\ln(\varphi^2)}{\ln(2)}(-2k\ln(\varphi))
}{-k \ln(2)} = -M[\frac{\ln(\varphi^2)}{\ln(2)}]^2,$$ which is,
exactly, $\bar{f}_H'(\bar{\al}_{max})\bar{\al}_{max}^2;$ i.e. the
RHS of (\ref{siete}).

\item{ 3) } We want now the value of $\al$ for which $q=0$. Let us
call $\beta$ that particular value of $\al$ for which
$q=\bar{f}_H'(\al) = 0.$
We have $\bar{f}_H(\beta) = 1$. On the other hand $f_H(\al) = \al
\bar{f}_H(\frac{1}{\al})$, so $f_H'(\al) =
\bar{f}_H(\frac{1}{\al}) - \frac{1}{\al}
\bar{f}_H'(\frac{1}{\al})$. Then, for $\frac{1}{\al} = \beta $ we
have $f_H'(\frac{1}{\beta}) = \bar{f}_H(\beta) - \beta
\bar{f}_H'(\beta) = 1$. Since the line $y  =x $ is tangent to $f_H(\al)=
f_C(\al)$ we have that $f_H(\frac{1}{\beta}) = \frac{1}{\beta}$.

Let us go back to $\bar{f}_H(\al)$. Our value $\beta$ is
obtained from

\begin{equation}
\left\{
\begin{array}{lcl}
q = \bar{f}_H'(\beta) = 0 \\
\bar{\tau}'(q) = \bar{\tau}'(0) = \beta
\end{array}
\right.
\end{equation}

Now, $\bar{\tau}'(q) = \lim_{k \to \infty} \frac{1}{ \sum^{2k}_{i=1}
p_i^q}\frac{ \sum p_i^q \ln(p_i)}{-k\ln(2)}$, which, for $q=0$
yields $\bar{\tau}'(0) = \beta = \frac{2 \bar{\pi}}{\ln(2)}$. Therefore,
the LHS of $\lim_{k \to \infty} \frac{\ln(\sum_{i=1}^{2^k}
p_i^q )}{-k\ln(2)} = \al q - \bar{f}_H(\al)$ |when $\al = \beta$,
$q= \bar{f}_H'(\beta) = 0$, and $\bar{f}_H(\al) = \bar{f}_H(\beta)
= 1$| is $\frac{\ln(2^k)}{-k \ln(2)} = -1$, whereas the RHS
is $\beta . 0 - \bar{f}_H(\beta) = -1 \ldots $so again we have
strict equality.

\subsection{The derivative of equation (\ref{siete})}

\medskip
We differentiate now both sides of equation (\ref{siete}) |$\al$ is our
variable| and we obtain $$\frac{1}{\sum^{2^k}_{i=1} p_i^{\phi + \al
\phi'}} \frac{\sum p_i^{\phi + \al \phi'}\ln(p_i)}{-k\ln(2)} (\phi
+\al \phi')' =\!\!\!?\ 2\al\phi'+ \al^2\phi''$$ $$= \al(2\phi'+
\al\phi")= \al(\phi + \al\phi')'.$$

Notice that, we can |due to the nature of $\phi$| safely cancel
$(\phi + \al\phi')'$, except for $\al =0$ and some other
occasional value of $\al$ in $(0,\bar{\al}_{max}].$ If we {\it do} so,
we are left with the interrogation $\bar{\tau}'(q)=\!\!\!?\ \al$ for $q =
\bar{f}_H'(\al).$ We use the values of $\al_n, \phi(\al_n)$, and
$\phi'(\al_n)$ in Table 4, in order to obtain Table 6,

\medskip

%Tabla 6
\begin{center}

TABLE 6: The values $\al_n$ and $\bar{\tau}'(\bar{f}_H'(\al_n))$.

\smallskip

\begin{tabular}{|c|c|c|c|c|c|c|}
\hline
  n                                          & 1      & 2      & 3      & 4      & 5      & 6
  \\ \hline
  $\bar{\tau}'(\bar{f}_H'(\al_n)) = {\bf X}(\al_n)$ & 0.7439 & 0.7749 & 0.7906 & 0.8332 & 0.8847 & 0.9719
  \\ \hline
  $\al_n = {\bf Y}(\al_n)$                     & 0.5423 & 0.6182 & 0.7016 & 0.7941 & 0.8696 & 0.9686  \\ \hline
\end{tabular}

\bigskip
\begin{tabular}{|c|c|c|c|c|c|c|c|}
\hline

 7 &   8      & 9      & 10     & 11     & 12     & 13     & 14
\\ \hline

 1.0726  &  1.14   & 1.2223 & 1.3007 & 1.3474 & 1.3696 & 1.3742 &
1.3753 \\ \hline

  1.0727 & 1.1431 & 1.2199 & 1.2972 & 1.3366 & 1.3548 & 1.3611 & 1.3642 \\ \hline
\end{tabular}

\end{center}

\medskip

and the corresponding relative errors $\Delta_n$, calculated by
excess as before are in Table 7. Again the curve $({\bf X}(\al_n),
{\bf Y}(\al_n))$ is indistinguishable from the line ${\bf Y}={\bf X}$.

\medskip

%Tabla 7
\begin{center}

TABLE 7: The errors $\Delta_n$ corresponding to the values
shown in Table 6. 

Again, we deal separately with values $n=1,\ 2,\ 3$ (for which $\Delta_n$ is big).

\smallskip

\begin{tabular}{|c|c|c|c|c|c|c|c|}
\hline
  $n$        & 1      & 2      & 3      & 4      & 5      & 6      & 7      \\
\hline
  $\Delta_n$ & 0.3718 & 0.2534 & 0.1269 & 0.0492 & 0.0174 & 0.0034 & -0.001 \\ \hline
\end{tabular}

\bigskip

\begin{tabular}{|c|c|c|c|c|c|c|}
\hline
  8       & 9      & 10     & 11     & 12     & 13     & 14      \\
\hline
  -0.0027 & 0.0020 & 0.0027 & 0.0081 & 0.0109 & 0.0096 & 0.0082  \\ \hline
\end{tabular}

\end{center}

\medskip

So we can safely conclude that equality $\bar{\tau}'(q) = \al$ for $q =
\bar{f}_H'(\al)$ holds.

\subsection{Theoretical aspects of equality $\bar{\tau}'(q) = \al$}

\medskip

\item{1)} As before, let us choose $\al = \bar{\al}_{max} =
\frac{\ln(\varphi^2)}{\ln(2)}$, hence  $\phi(\bar{\al}_{max}) = 0$
and $\phi'$ is $-M$, with the corresponding selection of $p_i =
p_i^{(k)} \cong \frac{1}{\varphi^{2k}}$. We have then $$\bar{\tau}'(0 +
\frac{\ln(\varphi^2)}{\ln(2)}(-M))= \frac{1}{(\frac{1}{\varphi^{2k}})^{0
+ \frac{\ln(\varphi^2)}{\ln(2)}(-M)}} \frac{(\frac{1}{\varphi^{2k}})^{0
+ \frac{\ln(\varphi^2)}{\ln(2)}(-M)} \ln(\frac{1}{\varphi^{2k}})}{-k
\ln(2)}$$

$$ = \frac{\ln(\frac{1}{\varphi^{2k}})}{-k \ln(2)} =
\frac{\ln(\varphi^2)}{\ln(2)} = \bar{\al}_{max}$$ which makes
$\bar{\tau}'(q) = \al$ for $q = \bar{f}_H'(\al)$ analytically correct.

\item{2)} Again, let us try $\al  = 0$, for which $\phi$ is unity and $\phi'$
zero. Then the exponent $q$ is unity, and for $\bar{\tau}_k'(q)$, i.e.
the expression for $\bar{\tau}'(q)$ {\it minus} $"\lim_{k \to
\infty}",$ we have: $\bar{\tau}_k'(q) = \bar{\tau}_k'(1) =
\frac{1}{\sum^{2^k}_{i=1} p_i } \frac{\sum^{2^k}_{i=1} p_i
\ln(p_i)}{-k \ln(2)} = -\frac{1}{\ln(2)} \frac{\sum^{2^k}_{i=1}
p_i \ln(p_i)}{k}$. We want to understand why
$\frac{\sum^{2^k}_{i=1} p_i \ln(p_i)}{k} \to 0 = \al $ as $k \to
\infty$, a fact that we can observe in the screen of our computer.
Again, we work with $k$ so large as to ensure that, for a certain
percentage $p$, sensibly larger than zero, we have $\ln(Q^{(k)}(i))
\approx \bar{\pi}_k \cong \bar{\pi}$. The tighter this resemblance,
the smaller $p$ will be. Then we have

$$ \frac{\sum^{2^k}_{i=1} p_i \ln(p_i)}{k} = \frac{
-\sum^{2^k}_{i=1} \frac{ \ln(Q_i^{(k)})}{Q_i^{(k)}Q_{i+1}^{(k)}} -
\sum^{2^k}_{i=1} \frac{\ln(Q_{i+1}^{(k)})}{Q_i^{(k)}
Q_{i+1}^{(k)}} }{k}$$ which is of the order of $2(\sum^{2^k}_{i=1
} \frac{\ln(Q_i^{(k)})}{Q_i^{(k)} Q_{i+1}^{(k)}})\frac{1}{k},$
 whose order is given by
$2^k\frac{\ln(e^{\bar{\pi}_k k})}{e^{\bar{\pi}_k k}e^{\bar{\pi}_k
k}} \frac{1}{k}.$

This value is $\bar{\pi}_k [\frac{2}{e^{2\bar{\pi}_k}}]^k$;
$\bar{\pi}_k$ is a stable multiplicative constant; it remains to
see if $\frac{2}{e^{2 \bar{\pi}_k}} < 1$ if $k \to \infty$, i.e.
if $2 < e^{2\bar{\pi}_k}$ or if $\ln(2) < 2\bar{\pi}_k$ $ \ldots$but
at the end of Section 4.2 we saw that $\frac{2\bar{\pi}_k}{\ln(2)}
> 1$ if $k$ is large |for $k = 22$ we had obtained $\frac{
2\bar{\pi}_k}{\ln(2)} = 1.1314$, and $\bar{\pi}$ was very slowly
increasing as $k$ grew.  So, again, for $\al = 0$ we have
$\bar{\tau}'(q) = \al$.

\item{3) } Finally, let us consider the value of $\al$ for which
$q=0$, i.e. $\al = \frac{2\bar{\pi}}{\ln(2)}.$ Therefore,
$\bar{\tau}'(q) = \lim_{k \to \infty} \bar{\tau}_k'(q) = \lim_{k \to \infty}
\frac{1}{\sum p_i^q} \frac{\sum p_i^q\ln(p_i)}{-k\ln(2)}$ becomes
$\lim_{k \to \infty} \frac{1}{2^k} \frac{\sum_{i=1}^{2^k}
\ln(p_i)}{-k\ln(2)} = $ $\lim_{k \to \infty} \frac{1}{\ln(2)}\frac{
\sum_{i=1}^{2^k} \frac{\ln(Q_i^{(k)} Q_{i+1}^{(k)})}{k}}{2^k} =
\frac{2 \bar{\pi}}{\ln(2)} = \al$, so again $\bar{\tau}'(q) = \al$ in a
strict way, when $\al = \frac{2 \bar{\pi}}{\ln(2)}$.

\section{The function $\phi(\al)$. Its Legendre Transformation}

\medskip

Let us consider $\phi(\al) = f_H(\frac{1}{\al}) =
d_H(\Omega_{\frac{1}{\al}})$, where

$$\Omega_{\frac{1}{\al}} = \{ x \in [0,1] / \forall k \in \N : x
\in I_i^{(k)}; \ l(I_i^{(k)}) = p_i^{(k)} =
\frac{1}{Q_i^{(k)}Q_{i+1}^{(k)}}; $$ $$ \mu(I_i^{(k)}) =
\frac{1}{2^k}; \  \mbox{and} \ \frac{\ln(1/2^k)}{\ln(p_i^{(k)})}
\to \frac{1}{\al} \ \mbox{when} \ k \to \infty \}.$$

In other words, $\phi(\al) = d_H( \{ x \in [0,1] / \
\frac{\ln(1/2^k)} {\ln(p_i^{(k)})} \to  \frac{1}{\al}; \  x \in
I_i^{(k)} \forall k \})$ for short. Here "$x \in I_i^{(k)}$"
means: $x$ belongs to a sequence of nested intervals of length
$p_i^{(k)} = \frac{1}{Q_i^{(k)}Q_{i+1}^{(k)}}.$ Notice that we can
write

\begin{equation}
\label{ocho} \phi(\al) = d_H( \{ x \in [0,1] /
\frac{\ln(p_i^{(k)})}{\ln(1/2^k)} \to\al \mbox{ when } \  k \to
\infty \}).
\end{equation}

Let us take some $x \sim RLRLRL\ldots $ in the F-B left-right
system in diagram G below. We trust the notation is obvious. At
the left of  diagram G we have our F-B system of segments of
length $p_i^{(k)}$ and $\mu$ measure $\frac{1}{2^k}$. At the right
the roles of length and $\mu$ measure are inverted, and we will
denote this with $\bar{p}$ for lengths and $\bar{\mu}$ for
measure. Notice that, in the $\bar{p}, \bar{\mu}$ system at the
right of the diagram we do have some element representative of the
spelling $RLRLRL\ldots$ in a sequence of nested intervals:

\smallskip

DIAGRAM G:

%DIAGRAM G
\setlength{\unitlength}{1cm}
\begin{picture}(12,11)

\put(2,-0.5){$\vdots$ etc.} \put(8,-0.5){$\vdots$ etc. }

\put(0,1){$\frac{1}{2} \hspace{.2cm} \line(1,0){4.5} \hspace{.2cm}
\frac{1}{1}$} \put(6,1){$\frac{1}{2} \hspace{.2cm} \line(1,0){4.5}
\hspace{.2cm} \frac{1}{1}$}

\put(1.7,.9){ $\underbrace{\hspace{0.1cm}}_{RLRLR} $}

\put(0,4){$\frac{1}{2} \hspace{.2cm} \line(1,0){4.5} \hspace{.2cm}
\frac{1}{1}$} \put(6,4){$\frac{1}{2} \hspace{.2cm} \line(1,0){4.5}
\hspace{.2cm} \frac{1}{1}$}

\put(1.5,3.9){ $\underbrace{\hspace{0.7cm}}_{RLRL} $}
\put(2.4,3.6){$\frac{5}{8}$}

\put(.5,4.2) {$p_{RLRL}^{(4)}=\frac{1}{5.8};\ \mu_{RLRL}^{(4)}=
\frac{1}{2^4}$}

\put(0,7){$\frac{1}{2} \hspace{.2cm} \line(1,0){4.5} \hspace{.2cm}
\frac{1}{1}$} \put(6,7){$\frac{1}{2} \hspace{.2cm} \line(1,0){4.5}
\hspace{.2cm} \frac{1}{1}$}

\put(1.5,6.9){ $\underbrace{\hspace{1.1cm}}_{RLR} $}
\put(1.2,6.6){$\frac{3}{5}$}

\put(.5,7.2) {$p_{RLR}^{(3)}=\frac{1}{5.3};\ \mu_{RLR}^{(3)}=
\frac{1}{2^3}$} \put(0,10){$\frac{1}{2} \hspace{.2cm}
\line(1,0){4.5} \hspace{.2cm} \frac{1}{1}$}
\put(6,10){$\frac{1}{2} \hspace{.2cm} \line(1,0){4.5}
\hspace{.2cm} \frac{1}{1}$}

\put(.5,9.9){ $\underbrace{\hspace{2.1cm}}_{RL} $} \put(2.6,9.5)
{$\frac{2}{3}$}

\put(.5,10.2) {$p_{RL}^{(2)}=\frac{1}{2.3};\ \mu_{RL}^{(2)}=
\frac{1}{2^2}$}

\put(7.7,.9){ $\underbrace{\hspace{0.1cm}}_{RLRLR} $}

\put(7.5,3.9){ $\underbrace{\hspace{0.7cm}}_{RLRL} $}
\put(8.4,3.6){$\frac{11}{16}$}

\put(6.5,4.2) {$p_{RLRL}^{(4)}=\frac{1}{2^4};\ \mu_{RLRL}^{(4)}=
\frac{1}{5.8}$}

\put(7.5,6.9){ $\underbrace{\hspace{1.1cm}}_{RLR} $}
\put(7.2,6.6){$\frac{5}{8}$}

\put(6.5,7.2) {$p_{RLR}^{(3)}=\frac{1}{2^3};\ \mu_{RLR}^{(3)}=
\frac{1}{5.3}$}

\put(6.3,9.9){ $\underbrace{\hspace{2.1cm}}_{RL} $} \put(8.6,9.5)
{$\frac{3}{4}$}

\put(6.5,10.2) {$p_{RL}^{(2)}=\frac{1}{2^2};\ \mu_{RL}^{(2)}=
\frac{1}{2.3}$}

\end{picture}

\smallskip

The spelling $R, RL, RLR, \ldots$ at the right of Diagram G gives
a sequence of nested intervals of length $\frac{1}{2^k}$. Now, if
one did not know better (if one did not know about the inverse
function), from (\ref{ocho}), from the nestedness of the intervals
of length $\frac{1}{2^k}$ and measure $p_i^{(k)}$, and from the
sameness of the grammar, one would be tempted to conjecture:

$$\phi(\al) = d_H(\bar{\Omega}_{\al}),$$
 where $\bar{\Omega}_{\al}$ refers  to the system at the right in diagram G.

{\it This is nonsense}, since, in the left system, the
element $$x \sim RLRLRL\ldots$$ is $\frac{\sqrt5 - 1}{2}$, i.e.
the number in the unit interval with the largest Markov
irrationality coefficient (that is, the "most irrational number"
in $[0,1]$), whereas the corresponding $RLRLRL\ldots$ in the right
system $\bar{p}, \bar{\mu}$ in diagram G is $\frac{2}{3}$ |a very
rational number! It means that $\Om_{\frac{1}{\al}}$ and
$\bar{\Om}_{\al}$ have little or nothing in common.

\bigskip

What would it take for these two sets to be alike |at least, so that
they would share  the same Hausdorff dimension? It would suffice, say, that the $\frac{1}{2^k}$-length segments and the
$p_i^{(k)}$-length segments were very much alike, i.e. that the
measures Euclidean and hyperbolic were very much alike;
{\it then}, we could conjecture $\phi(\al) =\!\!\!?\ 
d_H(\bar{\Omega}_{\al})$, where the corresponding measure is
indeed taken over a uniform partition ($\frac{1}{2^k}$ in the
$k^{th}$ step). {\it If} that were the case, we would have:

\begin{equation}
\label{nueve} \lim_{k \to \infty} \frac{\ln(\sum^{2^k}_{i=1}
[p_i^{(k)}]^{\phi'(\al)})}{\ln(\frac{1}{2^k})} = \al \phi'(\al) -
\phi(\al)
\end{equation}
and, if this equality was so sharp that we could differentiate both
sides of it |$\al$ our variable| and still obtain equality, we
would then have

\begin{equation}
\label{diez}
\bar{\tau}'(\phi'(\al)) = \al.
\end{equation}

\bigskip
The astonishing fact is that (\ref{nueve}) is true: again, the
LHS of (\ref{nueve}) is considered numerically, and the
RHS theoretically. Again the elements displayed in Table 8
are taken from Table 4. ${\bf X}(\al_n) = \frac{\ln(\sum_i
p_i^{\phi'(\al_n)})}{-k \ln(2)}; $ ${\bf Y}(\al_n) =
\al_n\phi'(\al_n) - \phi(\al_n):$

\medskip

%Tabla 8

\begin{center}

TABLE 8: ${\bf X}(\al_n) = \frac{\ln(\sum_i p_i^{\phi'(\al_n)})}{-k
\ln(2) };$ ${\bf Y}(\al_n) = \al_n\phi'(\al_n)-\phi(\al_n).$

\smallskip

\begin{tabular}{|c|c|c|c|c|c|c|c|}
\hline
  $n$        & 1       & 2       & 3       & 4       & 5       & 6      & 7       \\
\hline
  ${\bf X}(\al_n)$ & -1.0169 & -1.0703 & -1.0807 & -1.1315 & -1.1970 & -1.3360 & -1.6107 \\
\hline
  ${\bf Y}(\al_n)$ & -1.0057 & -1.0336 & -1.0395 & -1.0733 & -1.1207 & -1.23  & -1.4707 \\ \hline
\end{tabular}

\bigskip

\begin{tabular}{|c|c|c|c|c|c|c|}
\hline
  8       & 9       & 10      & 11       & 12       & 13       & 14       \\
\hline
  -1.9597 & -2.8786 & -5.5259 & -10.5229 & -17.8914 & -23.9972 & -29.2117 \\
\hline
  -1.7979 & -2.6760 & -5.3288 & -10.3522 & -17.7334 & -23.8149 & -29.0026 \\ \hline
\end{tabular}

\smallskip

\end{center}

\medskip

The plotting of the curve $({\bf X}(\al_n), {\bf Y}(\al_n))$ is in
Fig. 3:

\bigskip
%\begin{figure}
\begin{center}
\includegraphics[width=10cm,height=6cm]{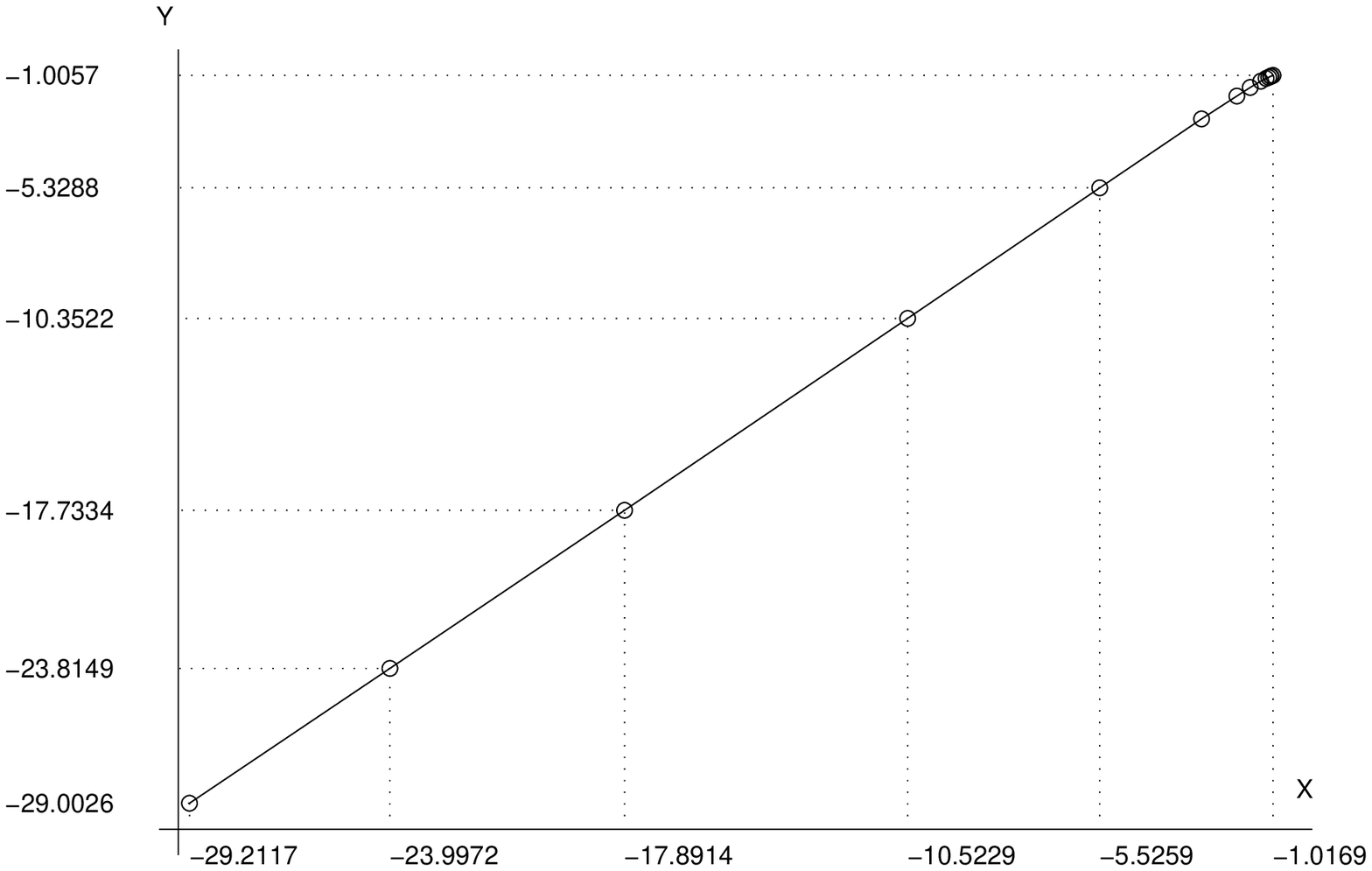}

%\caption{}
Fig. 3: The curve $({\bf X}(\al_n), {\bf Y}(\al_n));$ 
${\bf X}(\al_n)= \frac{\ln(\sum_ip_i^{\phi'(\al_n)})}{-k\ln(2)};
\  {\bf Y}(\al_n)=\al_n\phi'(\al_n)-\phi(\al_n). $
\end{center}

%\end{figure}

\medskip

\noindent a line indistinguishable from ${\bf Y}={\bf X}$.

The corresponding relative errors $\Delta_n$, estimated by excess
are shown in Table 9.

\medskip

%Tabla 9

\begin{center}

TABLE 9: The relative errors $\Delta_n$ corresponding to the
values in Table 8. 

\smallskip

\begin{tabular}{|c|c|c|c|c|c|c|}
\hline
  n          & 1      &    2    &      3  &      4  &      5  &     6   \\
\hline
  $\Delta_n$ & 0.0110 & -0.0343 & -0.0382 & -0.0510 & -0.0638 & -0.0793 \\
\hline
\end{tabular}

\bigskip
\begin{tabular}{|c|c|c|c|c|c|c|c|}
\hline
  7      & 8       & 9      & 10     & 11     & 12     & 13     & 14     \\
\hline
  0.0869 & -0.0825 & 0.0704 & 0.0357 & 0.0162 & 0.0088 & 0.0076 & 0.0072 \\ \hline
\end{tabular}

\end{center}

\medskip

{\it Moreover, if we differentiate both sides of equation (\ref{nueve}) we still obtain "equality"!}

Once we differentiate both sides of equation (\ref{nueve}), the
corresponding |tentative| equation now reads 
\begin{equation}
\label{once} \bar{\tau}'(\phi'(\al))\phi''(\al) =\!\!\!?\ \al\phi''(\al)
\end{equation}

The corresponding values |with $\phi''(\al_n)$ estimated discretely from our
values of $\al_n$ and $\phi'(\al_n)$ in Table 4| and the corresponding plot
$({\bf X}(\al_n), {\bf Y}(\al_n))$ are shown in Table 10 and Fig. 4.

\medskip

%Tabla 10

\begin{center}

TABLE 10: The RHS and LHS of (\ref{once}).

\smallskip

\begin{tabular}{|c|c|c|c|c|c|c|}
\hline
  n                              & 1     & 2    & 3    & 4    & 5    & 6
  \\ \hline
  ${\bf X}(\al_n)=$LHS of (\ref{once}) & -0.7  & -0.1 & -0.5 & -0.9 & -1.4 & -2.6
  \\\hline
  ${\bf Y}(\al_n)=$RHS of (\ref{once}) & -0.40 & -0.1 & -0.4 & -0.6 & -1.1 & -2.3 \\
  \hline
\end{tabular}

\bigskip
\begin{tabular}{|c|c|c|c|c|c|c|}
\hline
  7    & 8     & 9      & 10     & 11     & 12     & 13      \\
  \hline
  -5   & -12   & -34.3  & -126.9 & -404.2 & -975.6 & -1703.4 \\
  \hline
  -4.6 & -11.6 & -34    & -126.9 & -403.1 & -970.8 & -1693.9 \\ \hline
\end{tabular}

\end{center}
\medskip

%\begin{figure}
\begin{center}
\includegraphics[width=10cm,height=6cm]{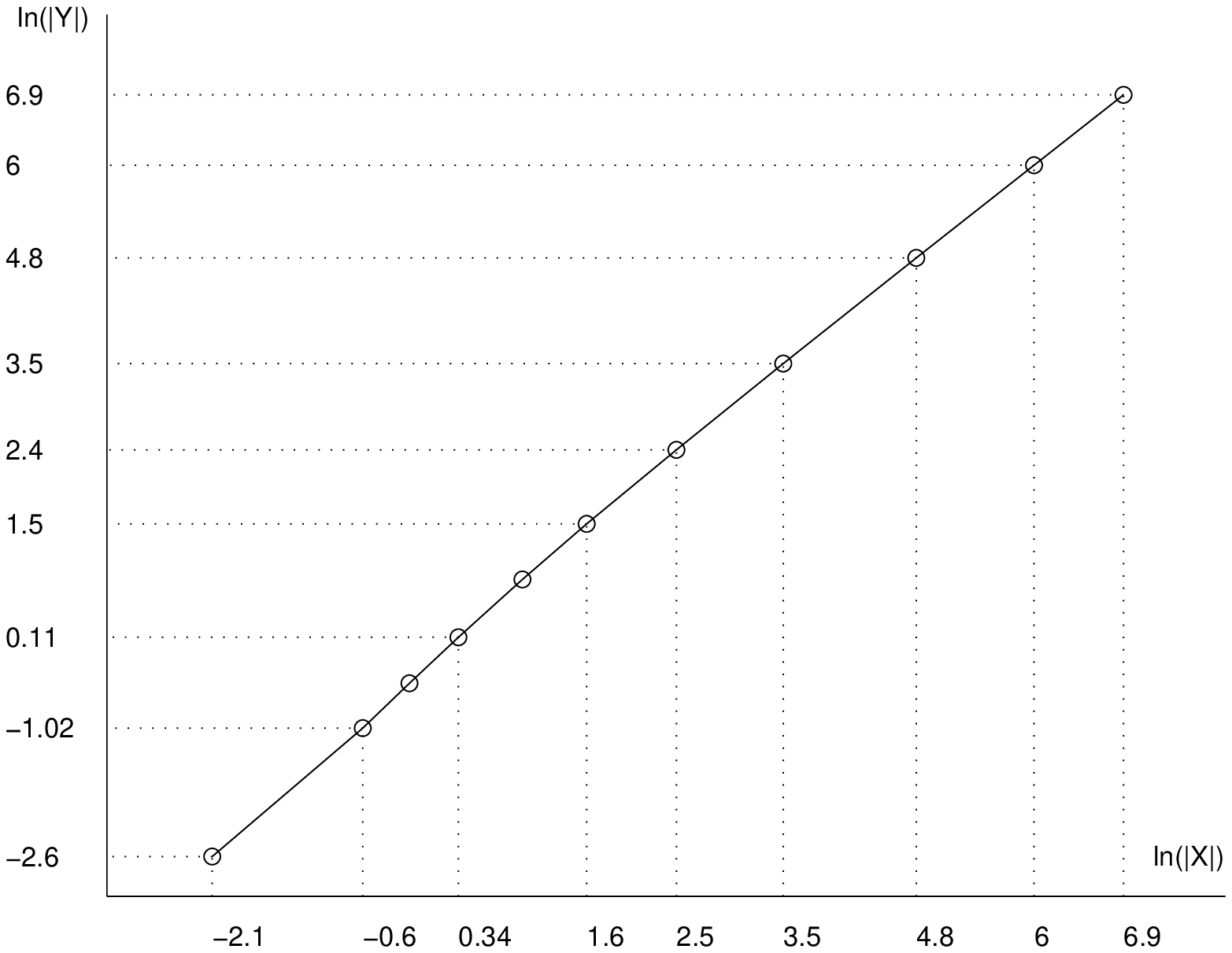}

%\caption{}

\smallskip

Fig. 4: The curve $(\ln(|{\bf X}|)(\al_n), \ln(|{\bf Y}(\al_n)|); $
${\bf X}(\al_n)=$LHS of (\ref{once}); ${\bf Y}(\al_n)=$RHS of (\ref{once}).
The points at the left of the curve correspond to small values of $\alpha$,
and will be dealt with in Section 6.1.
\end{center}
%\end{figure}

\bigskip

\bigskip
Notice that we did {\it not} cancel $\phi''(\al$) in both
sides of (\ref{once}), so that we are not really testing the
(impossible) truthfulness of (\ref{diez}).

\bigskip

Before we go any further, let us explore the

\subsection{Theoretical aspects of equations (\ref{nueve}),
(\ref{once}) and (\ref{diez})}

\medskip
Let us consider first equation (\ref{nueve})

\item{1)} for $\al = 0$. Then $\phi'(\al) =0$ and $\phi(\al) = 1.$
Clearly $\bar{\tau}(\phi'(\al)) = \bar{\tau}(0) = \lim_{k \to \infty}
\frac{\sum^{2^k}_{i=1} p_i^0}{\ln(\frac{1}{2^k})} = \lim_{k \to
\infty} \frac{\ln(2^k)}{-\ln(2^k)} = -1$, whereas $\al\phi'(\al) -
\phi(\al)$ is reduced to $-1$ as well.

\item{ 2)} Let us pose $\al = \frac{\ln(\varphi^2)}{\ln(2)}$. Then
$\phi(\al)=0$ and $\phi'(\al)=-M$, $M$ a huge number. As before,
the only surviving term in $\sum_{i=1}^{2^k} p_i^q$ is the one
corresponding to the smallest possible $p_i^{(k)}$, i.e.
$\frac{1}{\varphi^{2k}}$ times a certain constant $c$ |$k$ has to be
large. Therefore, $$\bar{\tau}(\al) = \lim_{k \to \infty}
\frac{\ln((\frac{1}{\varphi^{2k}})^{-M})}{-k\ln(2)} = (-M) \lim_{k \to
\infty} \frac{- 2k\ln(\varphi)}{-k\ln(2)} $$ $$ =
\frac{\ln(\varphi^2)}{\ln(2)}(-M) = \al \phi'(\al) - \phi(\al), $$
since $\phi(\al) = 0$.

Therefore, for the initial and final value of the domain of
$\phi$, Eq. (\ref{nueve}) is tight.

\bigskip

Let us consider equation (\ref{once})

\item{ 1)} for $\al = 0$. Now $\phi'(\al) = \phi''(\al) = 0.$ Let
us calculate $$\bar{\tau}'(\phi'(\al))  = \bar{ \tau}'(0)  =  \lim_{k \to
\infty} \frac{\sum^{2^k}_{i=1} p_i^0 \ln(p_i)}{ \sum_{i=1}^{2^k}
p_i^0 (-k \ln(2)) }$$ $$=  \lim_{k \to \infty}
\frac{\sum^{2^k}_{i=1} \ln(p_i)}{ 2^k (-k \ln(2)) }  = \lim_{k \to
\infty}- \frac{1}{\ln(2)} \frac{\sum^{2^k}_{i=1} \ln(p_i) /k
}{2^k}  $$ $$=  \frac{1}{\ln(2)} \lim_{k \to \infty} \frac{
(\sum^{2^k}_{i=1} \ln(Q_{(i)}^{(k)}) + \sum^{2^k}_{i=1} \ln(Q_{(i+1)}^{(k)})
)}{2^k} $$ $= \frac{1}{\ln(2)} \lim_{k \to \infty} 2 \bar{\pi}_k
= \frac{2 \bar{\pi}}{\ln(2)}, $ a finite number. The fact:
$\phi''(0) = 0,$ therefore, implies the validity of (\ref{once}) for $\alpha=0$. Notice that, if $\phi''(0)\not=0$, we would cancel it in equation (16), 
and we would be left with the interrogation

\begin{equation}
\bar{\tau}'(\phi'(\al)) =\!\!\!?\ \al  \ \mbox{ \ when \ } \ \al = 0.
\end{equation}

Notice that the answer is "no", for $\bar{\tau}'(0) =
\frac{2\bar{\pi}}{\ln(2)} \ne 0 = \al.$ Therefore $\bar{\tau}'(q) = \al$
for $q=f'(\al),$ i.e. (\ref{diez}) is false if $f=\phi$.

\item{ 2)} Let $\al= \frac{\ln(\varphi^2)}{\ln(2)}.$ Now, $\phi'' \ne
0$ and we can cancel it in both sides of equation (\ref{once}): we
are testing (\ref{diez}). We have now $\phi'(\al) = q = -M,$ $M$ a
huge value, and we know that, as before, sums $\sum_i p_i^q$ and
$\sum_ip_i^q\ln(p_i)$ are replaced by the term corresponding to
the smallest $p_i^{(k)}$ in the formula. We obtain, proceeding as
above,

\begin{equation}
\begin{array}{lcl}
\bar{\tau}'(-M) = \lim_{k \to \infty}
\frac{1}{(c\frac{1}{\varphi^{2k}})^{-M}}
\frac{(c\frac{1}{\varphi^{2k}})^{-M} \ln( c
\frac{1}{\varphi^{2k}})}{-k\ln(2)} \\ =\lim_{k \to \infty}
\frac{\ln(\frac{1}{\varphi^{2k}}) }{-k\ln(2)}=
\frac{\ln(\varphi^2)}{\ln(2)} = \al.
\end{array}
\end{equation}

Eq. (\ref{diez}) for $f= \phi$ is, therefore, tight for
$\bar{\al}_{max}.$

\bigskip
\noindent{\bf The second Legendre Equation for $\phi(\al)$}

Notice
that equation (\ref{once}) is essentially another form of
(\ref{diez}): $\phi'(\al)$ is
$-[\frac{f_H'(\frac{1}{\al})}{\al^2}],$ i.e. the  quotient of two
monotonic functions; therefore $\phi''(\al)$ is quite free to
change signs a number of times \ldots yet (with perhaps the
exception of $\al = 0$) each of these occasions will be an
isolated value of $\al$. Nevertheless we did notice that, for $\al
= 0$, we have $\bar{\tau}'(\phi'(\al)) \ne \al$, but, as $\phi''(\al) =
0$ for $\al = 0$, then (\ref{once}) held, even if (\ref{diez}) did
not: the value of $\phi''$ acted as a "leveler" for $\al = 0$
|i.e. when $\bar{\tau}'(q)$ was different from $\al$, $\bar{\tau}'\phi''$ was
equal to $\al\phi''$. Now, from Table 10 and the graph in Fig. 4
corresponding to (\ref{once}), we can infer that $\phi''$ is the
"leveler" for every $\al$! \ldots a mystery that implies,
somehow, that although (\ref{diez}) is not true, a certain relaxed
form of it is still true.

From bits and pieces in the last few sections, the reader can put
together a rather lengthy and technical proof of equation
(\ref{diez}) for a value of $\al$ strictly smaller than
$\bar{\al}_{max}$ (for which equation (\ref{diez}) holds), i.e.
for $\al = \frac{2\bar{\pi}}{\ln(2)}$ |a value slightly larger
than unity. We can then conjecture that, for the segment
$[\frac{2\bar{\pi}}{\ln(2)}, \bar{\al}_{max}]$ we do have equation
(\ref{diez}). But, in fact, though this is not a big interval, its
left extreme is difficult to reach ($k$ should be very large).
Therefore, we take a slightly smaller interval, and, for the
largest value of $k$ our fastest computer can handle we write:
${\bf X}(\al_n) = \bar{\tau}'(\phi'(\al_n))$; ${\bf Y}(\al_n) = \al_n$,
and we obtain Table 11.

\bigskip

%Tabla 11
\begin{center}

TABLE 11: ${\bf X}(\al_n) =\bar{\tau}'(\phi'(\al_n));$ ${\bf Y}(\al_n)
=\al_n.$

\smallskip

\begin{tabular}{|c|c|c|c|c|c|c|c|}
\hline
  $n$                    & 1      & 2      & 3      & 4      & 5      & 6      & 7      \\
\hline
  $\bar{\tau}'(\phi'(\al_n))$ & 1.2047 & 1.2410 & 1.2936 & 1.3352 & 1.3617 & 1.3709 & 1.3743 \\
\hline
  $\al_n$               & 1.1431 & 1.2199 & 1.2972 & 1.3366 & 1.3548 & 1.3611 & 1.3649 \\ \hline
\end{tabular}

\smallskip

\end{center}

\bigskip

The relative errors, computed, as usual, by excess, are displayed
in Table 12.

\bigskip

\begin{center}

TABLE 12: The relative errors $\Delta_n$ corresponding to the
values shown in Table 11.

\smallskip

%Tabla 12
\begin{tabular}{|c|c|c|c|c|c|c|c|}
\hline
  $n$        & 1      & 2      & 3       & 4       & 5      & 6      & 7      \\
\hline
  $\Delta n$ & 0.0538 & 0.0173 & -0.0028 & -0.0010 & 0.0051 & 0.0072 & 0.0074 \\ \hline
\end{tabular}

\end{center}

\medskip

These facts tell us that a segment contained in the line $ {\bf Y}={\bf X}$
is contained in the graph of $(\bf{X}(\al), \bf{Y}(\al))$,
$\bf{X}(\al) = \bar{\tau}'(\phi'(\al)); \ \bf{Y}(\al) = \al.$ In fact,
plotting such graph from values taken from all tables above,
we do notice a smooth curve, smoothly glued to a segment
contained in ${\bf Y}={\bf X}$,  $\al \in [\frac{2\bar{\pi}}{\ln(2)},
\bar{\al}_{max}].$ It means that (\ref{diez}), without $\phi''$ as
a "leveler", holds in a substantial part of the domain of $\phi$.

\section{Conclusion} 

The mystery of $\phi(\al)$ posed by
equations (\ref{nueve}) and (\ref{diez}) can be explained by
conjecturing that the Lebesgue and the hyperbolic measures are not
so terribly different as they seem. Another way of saying this:
$[0,1],$ as a fractal with $\mu^{\H}$ as a probability measure,
is more self-similar than it seems. Yet another way: $\mu^{\H}$ has
much more in common with the so called "self-similar measures"
than what it looks like. In order to clarify this, let us consider
certain examples of fractals contained in $[0,1]$, given as the
limit of an iterative process. The step of the iterations is
given, as usual, by $k \in \N$. 
The $k^{th}$ iteration of the
process is associated with a $k$-partition covering the fractal.
We will assume that all segments in the $k$-partition are
equiprobable.

Let us consider, first, the usual ternary Cantor set  as
$\Omega$. The $k^{th}$ partition consists of $2^k$ segments.  Their
length is $\frac{1}{3^k}$. Their reciprocal length is $3^k$: an
exponential with 3 as a base. The average $\bar{\pi}_k$ of all
the logarithms of these bases is $\ln(3)$, and so is the limit
$\bar{\pi}$ of $\bar{\pi}_k$ as $k \to \infty$.

Let us now consider $\Om$ as the unit segment; the corresponding
measure gives equiprobability to the two segments $[0,a]$ and
$[a,1]$, when $a \in (0,1)$, $a \ne 1/2$, $a$ can be irrational.
We have two contractions with factors $a$  and $b = 1 - a$
involved. When $k=2$ we will have four equiprobable segments of
lengths $a^2, ab, ba$ and $b^2, \ldots$ and so on. In the $k^{th}$
equiprobable iteration of the process we have $2^k$ segments of
length $a^ib^{ k-i}, \ 0 \le i \le k.$ If we express $\frac{1}{a^ib^{k-i}}$
as an exponential with exponent $k$, then the logarithm of the
base of the  exponential, $\ln(Q_{(i)}^{(k)}),$ will be a number
obviously bigger than $\frac{1}{2}\min \{ \ln(\frac{1}{a}), \ln(\frac{1}{b})\} = A$
and smaller than $\ln(\frac{1}{a}) + \ln(\frac{1}{b}) = B$.

The corresponding average $\bar{\pi}_k,$ therefore, will be in 
 interval $(A,B)$, and so will be its limit
$\bar{\pi}$ |which can be proved to exist.

We can continue in this way and convince ourselves that, given any
finite number of contractions, the corresponding average limit
$\bar{\pi} = \bar{\pi}(\Omega)$ will exist.

Now, when we look at the $k^{th}$ F-B partition we find segments
whose lengths vary from $\frac{1}{k}$ to
$\frac{1}{F^{(k)}F^{(k+1)}},$ where $F^{(k)}$ is the $k^{th}$
Fibonacci number, of the order of $\varphi^k.$ If we consider, for
simplicity, the reciprocal lengths, we find that they go from $k$ to
$\varphi^{2k},$ passing through, say, $3k^2, 5k^3, \ldots$ you name
it: $\Omega$ looks strongly non self-similar, it looks the opposite of
having a finite number of contractors$\ldots$ Yet, the
corresponding $\bar{\pi}_k$ grows with $k$ and reaches a value
$\bar{\pi}$ very much alike its theoretical maximum: it means
that segments of non-exponential reciprocal length, such as $k,
3k^2, \ldots$ etc., are, on the whole, negligible in number, and
$\mu^{\H}$ seems to behave, asymptotically, as a "self-similar
measure", despite all the irreconcilable differences between both
measures referred to in the Introduction.

\bigskip

\noindent{\bf References}

Arrowsmith, D.K., Lansbury, A., \& Mondrag\'on, R. [1996]
''Control of the Arnold Circle Map," {\it International Journal
of Bifurcation and Chaos} {\bf 6}(3), 437-453. 

Bak, P. [1986] ''The devil's staircase," {\it Physics Today }, Dec., 38-45.

Bruinsma, R. \& Bak, P. [1983] ''Self-similarity and fractal dimension of the devil's staircase in the one-dimensional Ising model," 
{\it Phys. Rev.} {\bf B  27}(9), 5924-5925.

Cawley, R. \& Mauldin, R. D. [1992] ''Multifractal decompositions of Moran
fractals,"
{\it Adv. in Math. } {\bf 92}(2), 196-236. 

Cesaratto, E. [1999] ''On the Hausdorff dimension of certain sets arising in Number Theory," {\it Los Alamos National Laboratory, xxx.lanl.gov}.

Cesaratto E., \& Piacquadio, M. [1998] ''Multifractal formalism of the Farey partition," {\it Revista de la Uni\'on
Matem\'atica Argentina,} {\bf 41}(2), 51-66. 

Cvitanovic, P., Jensen, H., Kadanoff, L. P. \&  Procaccia, I. [1985]
''Renormalization, unstable manifolds, and the fractal structure of mode locking,"
{\it Phys. Rev. Lett. } {\bf 55}(4),  343-346.

Duong-Van, M. [1987] ''Phase transition of multifractals," {\it Nuclear Physics B (Proc. Suppl.)} {\bf 2}, 521-526. 

Gruber, C., Ueltschi, D. \& Jedrzejewski, J. [1994]  ''Molecule formation and the Farey tree in the one-dimensional Falicov-Kimball model," {\it
Journal of Stat. Phys. } {\bf 76}(1/2), 125-157. 

Grynberg, S. \& Piacquadio, M. [1995] ''Hyperbolic Geometry and
Multifractal Spectra. Part II," {\it Trabajos de Matem\'aticas 252,
Publicaciones Previas del Instituto Argentino de Matem\'aticas, I.
A. M. -CONICET. }

Halsey, T. C., Jensen, M. H., Kadanoff, L. P., Procaccia, I. \&
  Schraiman, B. [1986] ''Fractal measures and their singularities: the characterization of strange sets," {\it Nucl. Phys. B, Proc. Suppl. 2} 513-516.

McGehee, R. P. \& Peckham, B. [1996] ''Arnold flames and resonances surface folds," {\it International Journal 
of Bifurcation and Chaos,} {\bf 6}(2), 315-336.

Mandelbrot, B. \& Riedi, R. [1997] ''Inverse measures, the inversion formula and discontinuous multifractals," {\it Adv. Appl. Math.} {\bf 18}, 50-58.

Piacquadio, M. \& Grynberg, S. [1998] ''Cantor staircases in Physics and diophantine approximations," {\it The International
Journal of Bifurcation and Chaos, } {\bf 8}(6), 1095-1106. 

Riedi, R. \& Mandelbrot, B. [1997] ''The inversion formula for continuous multifractals," {\it Adv. Appl. Math.} {\bf 19},
332-354.

Riedi, R. \& Mandelbrot, B. [1998] ''Exceptions to the multifractal formalism for discontinuous measures," {\it Math. Proc. Camb. Phil. Soc.}
{\bf 123}, 133-157.

Rosen, M. [1998] private communication

Series, C. [1985] ''The modular surface and continued fractions," {\it Journal of the London Mathematical
Society (2),} {\bf 31}(1), 69-80.

Series, C. \& Sinai, Y. [1990] ''Ising models on the Lobachevsky plane,"
{\it Communications in Math.
Physics,} {\bf 128}, 63-76. 

Tel, T. [1988] ''Fractals, multifractals and thermodynamics: an introductory review," {\it Zeitschrift fur Naturforschung, }
{\bf 43a}, 1154-1174.

\end{subsection}

\end{document}